\documentclass[onecolumn]{fairmeta}

\title{LIME: Link-based User-item Interaction Modeling with Decoupled XOR Attention for Efficient Test Time Scaling}
\PassOptionsToPackage{numbers}{natbib}
\usepackage[most]{tcolorbox}
\usepackage{amsmath}
\usepackage{amssymb}
\usepackage{enumitem}  
\usepackage{tabularx}
\usepackage[numbers]{natbib} 
\usepackage{booktabs} 
\usepackage{multirow}
\usepackage{pifont}   
\usepackage{xcolor}
\usepackage[utf8]{inputenc}
\usepackage[table]{xcolor} % 单元格着色
\usepackage{siunitx}    % 小数点精准对齐
\usepackage{amsfonts}   % 字体支持
\usepackage{caption}
\usepackage{makecell}

\usepackage[T1]{fontenc}    % use 8-bit T1 fonts
\usepackage{hyperref}       % hyperlinks
\usepackage{url}            % simple URL typesetting
\usepackage{booktabs}       % professional-quality tables
\usepackage{amsfonts}       % blackboard math symbols
\usepackage{nicefrac}       % compact symbols for 1/2, etc.
\usepackage{microtype}      % microtypography
\usepackage{newfloat}
\usepackage{listings}

\usepackage{subcaption}
\usepackage{amsmath}
\usepackage{array}
\usepackage{amsfonts}
\usepackage{booktabs}
\usepackage{comment}
\usepackage{siunitx}
\usepackage{algorithm}
\usepackage{algpseudocode}
\usepackage{amsmath}

\DeclareMathOperator{\tr}{\operatorname{Tr}}

\DeclareMathOperator{\XORA}{\operatorname{XORA}}
\DeclareMathOperator{\MHA}{\operatorname{MHA}}

\usepackage{tikz}
\usepackage{multirow}
\usepackage{pgfplots}

\usepackage{fontawesome5} % For the device icons
\usepackage{xcolor}

\usetikzlibrary{positioning, fit, shapes.geometric, arrows.meta, backgrounds, shapes.arrows}

\newcommand{\probP}{\text{I\kern-0.15em P}}
\let\oldforall\forall
\renewcommand{\forall}{\, \oldforall \, }

\let\oldexist\exists
\renewcommand{\exists}{\: \oldexist \: }

\definecolor{steelblue}{HTML}{4682B4}
\colorlet{boxcolor}{steelblue}  % Options: orange-web, sagegreen, lemonyellow, skyblue, coral, lavender, mintgreen, peach, steelblue, rosegold

\newtcolorbox{takeawaybox}[1][]{
  enhanced,
  attach boxed title to top left={xshift=4mm,yshift=-2mm},
  colback=boxcolor!10,
  colframe=boxcolor!60,
  colbacktitle=boxcolor!80,
  coltitle=white,
  fonttitle=\bfseries\small,
  boxed title style={size=small, colframe=boxcolor!80, sharp corners},
  sharp corners,
  boxrule=0.8pt,
  left=4pt, right=4pt, top=4pt, bottom=4pt,
  breakable,
  title={#1}
}

\author[\dagger*]{Yang Liu}
\author[*]{Yunjiang Jiang}
\author[*]{Ayush Agarwal}
\author{Yihan Wu}
\author{Haoran Liu}
\author{Bi Xue}

\affiliation{Meta AI}

\contribution[*]{Equal Contribution}
\contribution{\texttt{yangliu991@meta.com}}
\contribution{\texttt{yunjiang.jiang@fireworks.ai}}
\contribution{\texttt{yush94582@gmail.com}}
\contribution{\texttt{yihanwu@meta.com}}
\contribution{\texttt{ryan0814@meta.com}}
\contribution{\texttt{bixue@meta.com}}

\correspondence{Yang Liu}
\begin{document}

\abstract{
 Scaling large recommendation systems requires advancing three major frontiers: processing longer user histories, expanding candidate sets, and increasing model capacity. While promising, transformers' computational cost scales quadratically with the user sequence length and linearly with the number of candidates. This trade-off makes it prohibitively expensive to expand candidate sets or increase sequence length at inference, despite the significant performance improvements.
% \par\medskip\par
We introduce \textbf{LIME}, a novel architecture that resolves this trade-off. Through two key innovations, LIME fundamentally reduces computational complexity. First,  low-rank ``link embeddings" enable pre-computation of attention weights by decoupling user and candidate interactions, making the inference cost nearly independent of candidate set size. Second, a linear attention mechanism, \textbf{LIME-XOR}, reduces the complexity with respect to user sequence length from quadratic ($O(N^2)$) to linear ($O(N)$).
% \par\medskip\par
Experiments on public and industrial datasets show LIME achieves near-parity with state-of-the-art transformers but with a 10$\times$ inference speedup on large candidate sets or long sequence lengths. When tested on a major recommendation platform, LIME improved user engagement while maintaining minimal inference costs with respect to candidate set size and user history length, establishing a new paradigm for efficient and expressive recommendation systems.
}
\maketitle

\date{\today}
\correspondence{First Author at \email{first.author@meta.com}}

% You can add additional metadata fields as follows 
\metadata[Code]{\url{https://github.com/facebookresearch/repo}}
\metadata[Blogpost]{\url{https://ai.meta.com/blog/?page=1}}

\begin{figure*}[!htbp]
    \centering
    % --- FIRST PLOT: vs. Number of Candidates ---
    \begin{subfigure}[b]{0.33\textwidth}
        \centering
        \begin{tikzpicture}
            \pgfplotstableread{
                max_nro_len  lime_infer   mha      pt_stu_sky  pt_lime_xor_infer
                16.0   0.838037   0.868014   3.598747   2.616014
                32.0   0.831868   0.868626   3.659773   2.531268
                64.0   0.877271   0.878609   3.742255   2.515830
                128.0   0.844882   0.870173   3.635680   2.548429
                256.0   0.871572   0.880626   3.669382   2.517248
                512.0   0.852563   0.884552   4.013297   2.557467
                1024.0   0.869424   0.891254   3.764502   2.606724
                2048.0 0.852571   0.947399   3.699092   2.426673
                4096.0   0.829745   1.055303   4.170820   2.513015
                8192.0   0.856446   1.112514   4.416213   2.528355
                16384.0   0.873041   1.642221   7.862438   3.212758
                32768.0   0.987450   2.762807   11.662804   4.057047
                65536.0   1.351057   4.251611   23.831122   5.947808
            }\datatable
            \begin{semilogxaxis}[
                xlabel={Number of Candidates},
                ylabel={Latency (ms)},
                grid=both, grid style={dashed, gray!30},
                width=\linewidth, height=5cm,
                log basis x=2,
                xtick={64, 256, 1024, 4096, 16384, 65536},
                xticklabel style={rotate=45, anchor=east},
                % removed invalid legend show key
            ]
                \addplot[color=blue, mark=*] table [x=max_nro_len, y=lime_infer] {\datatable};
                \addplot[color=red, mark=x] table [x=max_nro_len, y=mha] {\datatable};
                \addplot[color=green!50!black, mark=triangle*] table [x=max_nro_len, y=pt_stu_sky] {\datatable};
                \addplot[color=orange, mark=diamond*] table [x=max_nro_len, y=pt_lime_xor_infer] {\datatable};
            \end{semilogxaxis}
        \end{tikzpicture}
    \end{subfigure}%
    % \hfill
    % --- SECOND PLOT: vs. User History Length ---
    \begin{subfigure}[b]{0.33\textwidth}
        \centering
        \begin{tikzpicture}
            \pgfplotstableread{
                max_uih_len  lime_infer   mha      pt_stu_sky   pt_lime_xor_infer
                16.0    0.681039   0.715669            4.177987                  3.024185
                32.0    0.759579   0.800573            3.967658                  2.995801
                64.0    0.830852   0.716231            4.191226                  3.013176
                128.0    0.709616   0.714126            5.747041                  2.967373
                256.0    0.716385   0.701695            3.998695                  3.247496
                512.0    0.814207   0.730676            4.171691                  3.287557
                1024.0    0.774501   0.849275            4.111623                  2.935691
                2048.0    0.682174   0.914787            4.831791                  3.053527
                4096.0    0.873759   1.633177            7.830352                  3.036117
                8192.0    0.882934   1.616990            7.721833                  3.055694
                16384.0    1.393705   3.555888           14.003201                  3.842386
                32768.0    2.542677   7.288437           32.776440                  6.253336
            }\datatable
            \begin{semilogxaxis}[
                xlabel={User History Length},
                grid=both, grid style={dashed, gray!30},
                width=\linewidth, height=5cm,
                log basis x=2,
                xtick={64, 256, 1024, 4096, 16384},
                xticklabel style={rotate=45, anchor=east},
            ]
                \addplot[color=blue, mark=*] table [x=max_uih_len, y=lime_infer] {\datatable};
                \addplot[color=red, mark=x] table [x=max_uih_len, y=mha] {\datatable};
                \addplot[color=green!50!black, mark=triangle*] table [x=max_uih_len, y=pt_stu_sky] {\datatable};
                \addplot[color=orange, mark=diamond*] table [x=max_uih_len, y=pt_lime_xor_infer] {\datatable};
            \end{semilogxaxis}
        \end{tikzpicture}
    \end{subfigure}%
    % \hfill
    % --- THIRD PLOT: vs. QK Dimension ---
    \begin{subfigure}[b]{0.33\textwidth}
        \centering
        \begin{tikzpicture}
            \pgfplotstableread{
                D      lime_infer   mha      pt_stu_sky  pt_lime_xor_infer
                16.0   1.218069     1.283479   7.463579    2.563631
                32.0   1.284705     1.376948   7.099568    2.551324
                64.0   1.254182     1.309108   7.018779    2.574806
                128.0  1.258400     1.444284   7.205290    2.592434
                256.0  1.258016     1.881922   8.184027    2.654552
                512.0  1.238591     1.663589   7.543578    2.638407
                1024.0 1.301714     3.130324   11.021116   2.904856
                2048.0 1.506880     3.307019   10.777114   3.359295
                4096.0 3.065670     13.574147  35.229552   7.927708
            }\datatable
            \begin{semilogxaxis}[
                xlabel={QK dim},
                grid=both, grid style={dashed, gray!30},
                width=\linewidth, height=5cm,
                log basis x=2,
                xtick={16, 64, 256, 1024, 4096},
                xticklabel style={rotate=45, anchor=east},
            ]
                \addplot[color=blue, mark=*] table [x=D, y=lime_infer] {\datatable};
                \addplot[color=red, mark=x] table [x=D, y=mha] {\datatable};
                \addplot[color=green!50!black, mark=triangle*] table [x=D, y=pt_stu_sky] {\datatable};
                \addplot[color=orange, mark=diamond*] table [x=D, y=pt_lime_xor_infer] {\datatable};
            \end{semilogxaxis}
        \end{tikzpicture}
    \end{subfigure}

    % %\vspace{2mm}

    % --- SHARED LEGEND ---
    \begin{tikzpicture}
        \begin{axis}[
            hide axis,
            xmin=0, xmax=1, ymin=0, ymax=1,
            legend columns=-1,
            legend style={draw=none, column sep=0.5cm, anchor=north},
        ]
            \addlegendimage{color=red, mark=x}
            \addlegendentry{MHA}
            \addlegendimage{color=green!50!black, mark=triangle*}
            \addlegendentry{HSTU}
            \addlegendimage{color=blue, mark=*}
            \addlegendentry{LIME-MHA}
            \addlegendimage{color=orange, mark=diamond*}
            \addlegendentry{LIME-XOR}
        \end{axis}
    \end{tikzpicture}
    \vspace{-5cm}
    \caption{Overall latency analysis across different model parameters. This figure illustrates the inference latency vs number of candidates, user history length, and QK dimension of MHA, HSTU, LIME-MHA, and LIME-XOR. LIME-MHA consistently achieves the lowest and most stable latency across all three metrics.}
    \label{fig:main_latency_analysis}
    % %\vspace{-1cm}
\end{figure*}

\section{Introduction}
Modern recommendation systems operate at a massive scale, facing the challenge of ranking millions of candidate items within strict real-time latency constraints. To succeed, ranking models must navigate a fundamental trade-off between computational efficiency and predictive accuracy. This has led to two dominant but conflicting architectural paradigms. On one end of the spectrum is the two-tower model~\citep{covington2016deep}, which achieves unparalleled inference speed by encoding users and items into separate, independent representations. This separation enables the use of efficient approximate nearest-neighbor search~\citep{MIPS}, making real-time recommendations feasible at scale.

On the other end are powerful Transformer models like BST~\citep{chen2019behavior}, which deliver state-of-the-art accuracy by encoding users' long interaction history (UIH) via multi-layer self attention and explicitly modeling deep, contextual interactions between that and each candidate item. Although this approach provides rich expressiveness, it comes with a significant computational cost: the self-attention over long sequences and the cross-attention between the full user sequence and each candidate become major performance bottlenecks. This architectural dilemma is becoming increasingly acute. The push for higher quality recommendations now demands scaling along three axes simultaneously: accommodating vast candidate sets, processing longer user histories, and deploying models of increasing complexity. These demands are fundamentally at odds, as existing efficient models lack expressiveness, while expressive models lack efficiency. Hybrid approaches~\citep{li2022inttower} offer only incremental improvements.

To fundamentally solve this problem and enable scaling along all three axes, we propose a new architectural blueprint: the Link-based User-Item Interaction Modeling for Efficient inference (LIME) framework. LIME is designed from the ground up to achieve the modeling power of a full cross-attention system while operating within the strict efficiency budget of a two-tower model. Its central innovation is a globally learned intermediate "link embedding" sequence that acts as a bridge between the long user history and candidate items. This design decouples the user and item representations during online inference, making scoring independent of history length by pre-computing the most expensive attention components offline. Furthermore, it enables the introduction of a new low rank attention mechanism to reduce user interaction history (UIH) self-attention complexity from quadratic to linear. By resolving these two primary computational bottlenecks, LIME provides a comprehensive solution to the expressiveness-versus-efficiency challenge, enabling deep interaction modeling at minimal latency (Figure~\ref{fig:main_latency_analysis}).

Our primary contributions are as follows:
% \begin{itemize}
% \item \textbf{A Novel Architecture for Scalable Ranking:} We propose LIME, a framework that resolves the core tension between efficient two-tower and expressive cross-attention models. It directly addresses the need to simultaneously scale along three dimensions: candidate set size, user history length, and model complexity.

\textbf{The Link Embedding Mechanism:} We introduce a link embedding sequence that effectively approximates full cross-attention. This mechanism allows the expensive Query-Key attention weight, $\phi(\tilde{Q} \tilde{K}^\top)$, to be pre-computed and cached offline, enabling cross-attention-like expressiveness with the efficiency of Two-Tower modeling during online inference.

\textbf{XOR Attention Masking:} To overcome the quadratic time complexity of Transformers with respect to user history length, we propose an XOR attention mask that factorizes the full self-attention matrix into a bidirectional linear attention between the link embeddings and the user history sequence.

\textbf{State-of-the-Art Performance and Impact:} We demonstrate through extensive experiments that LIME achieves performance competitive with computationally intensive ranking models like HSTU~\citep{zhai2024actions} with 10x lower latency. Deployed in production, LIME has yielded up to 38\% source rate\footnote{This refers to the percentage of positively engaged items attributable to the ranking model using LIME.} gain on a major platform serving billions of users.
% \end{itemize}

\section{Related Work}
LIME addresses the long-standing trade-off between model expressiveness and inference efficiency in large-scale ranking. We situate our contributions in the context of two primary research areas: efficient ranking architectures and innovations in sequence modeling.

\subsection{Efficient Ranking Architectures}
The design of ranking models is dominated by a conflict between efficiency and interaction depth. On one end of the spectrum, two-tower models~\citep{covington2016deep, yi2019sampling} achieve unparalleled efficiency. By encoding users and items into separate embedding spaces, they enable fast candidate retrieval using Approximate Nearest Neighbor (ANN) search. However, this separation prevents deep, feature-level interactions, limiting model expressiveness. On the other end, cross-attention models like DIN~\citep{zhou2018deep} enable rich, target-aware interactions by dynamically attending to user history for each candidate, but their per-item computational cost makes them prohibitive for ranking large candidate sets.

Several approaches have sought to bridge this gap. Hybrid models add shallow interaction layers on top of a two-tower base~\citep{li2022inttower}, though these often provide only marginal improvements. LIME offers a more fundamental solution. It introduces a novel bridge built upon a fixed set of global, learnable parameters, which we term link embeddings. These link embeddings act as an intermediary, allowing LIME to capture the rich dynamics of cross-attention while retaining the architectural efficiency of a two-tower model. This design, where a static, user-independent key space (raw links) facilitates the retrieval of information from a dynamic, personalized value space (personalized links), is conceptually similar to key-value memory networks~\citep{miller2016key} and the retrieval stage of retrieval-augmented models, enabling massive pre-computation. This approach directly tackles the sequential nature of user history and the need for scalable target-item interaction, striking a new balance between efficiency and expressiveness.

\subsection{Efficient Attention for Long Sequence Modeling}
The Transformer's quadratic complexity ($O(N^2)$) for self-attention remains a fundamental bottleneck for modeling long sequences. This challenge has driven the recent advances in Large Language Models (LLMs), moving beyond early approximations such as kernelization~\citep{choromanski2020rethinking} or low-rank projections~\citep{wang2020linformer}. More recent breakthroughs include new model classes—State Space Models (e.g., Mamba~\citep{gu2024mamba})—and hardware-aware techniques such as Lightning Attention~\citep{qin2024lightning} that directly optimize attention computation.

Within recommender systems, progress has often focused on adapting NLP-inspired attention mechanisms to click-through rate (CTR) prediction~\citep{zhang2022fum, chen2022efficient, Song_2025} or using pruning to shorten sequences before attention is applied~\citep{pi2020search}. While general-purpose mechanisms for handling set-based inputs exist, they often do not align with the specific needs of sequential recommendation. For instance, the Set Transformer~\citep{lee2019set} and its Pooling Multi-Head Attention (PMA) mechanism efficiently summarize a set into a fixed-size representation using learnable seed vectors. However, their primary goal is permutation-invariant summarization, whereas ranking requires target-aware representations that preserve the sequential nature of user histories.

LIME's XOR Attention contributes a distinct, task-specific solution. Rather than being a general-purpose approximation of the self-attention matrix, it is a mechanism co-designed with the LIME architecture. By using link embeddings as intermediaries, LIME structurally eliminates the need for direct history-to-history self-attention at inference time. This design enforces a linear complexity ($O(\ell \cdot N)$) tailored specifically for the user–item ranking context, representing a novel approach to building efficient and expressive sequence models for recommendations.
% %\vspace{-2em}
\section{Model Overview} \label{sec:model-overview}

We propose the \textbf{L}ink-based user-item \textbf{I}nteraction \textbf{M}odeling for \textbf{E}fficient inference (\textbf{LIME}), a novel sequential-modeling architecture tailored for Click-Through Rate (CTR) prediction. LIME is designed to bridge the gap between highly efficient but less expressive two-tower models and powerful but computationally expensive cross-attention architectures. We present its design by progressively building from a simple, efficient baseline to a scalable model with deeper interactions.

To represent a user $U$'s interaction history, we learn embedding tables to generate embeddings for each of the $N(U)$ items the user has interacted with. Each item is characterized by a set of attributes, which can be categorical (e.g., user action, topic id) or continuous (e.g., video length), and we learn embedding tables for each attribute. Continuous features are first transformed into categorical ones via bucketization to index into the embedding tables. For each item, we concatenate the learned embeddings of all its attributes and project them through a Multi-Layer Perceptron (MLP) to obtain a unified representation, $E_j$. The entire user history is then represented as a sequence of these embeddings, $E=\{E_j\}_{j=1}^{N(U)} \in \mathbb{R}^{N(U) \times d}$.

\begin{figure*}[t]
% %\vspace{-2em} % Reduce the space below this figure by 15 points
\includegraphics[width=\textwidth]{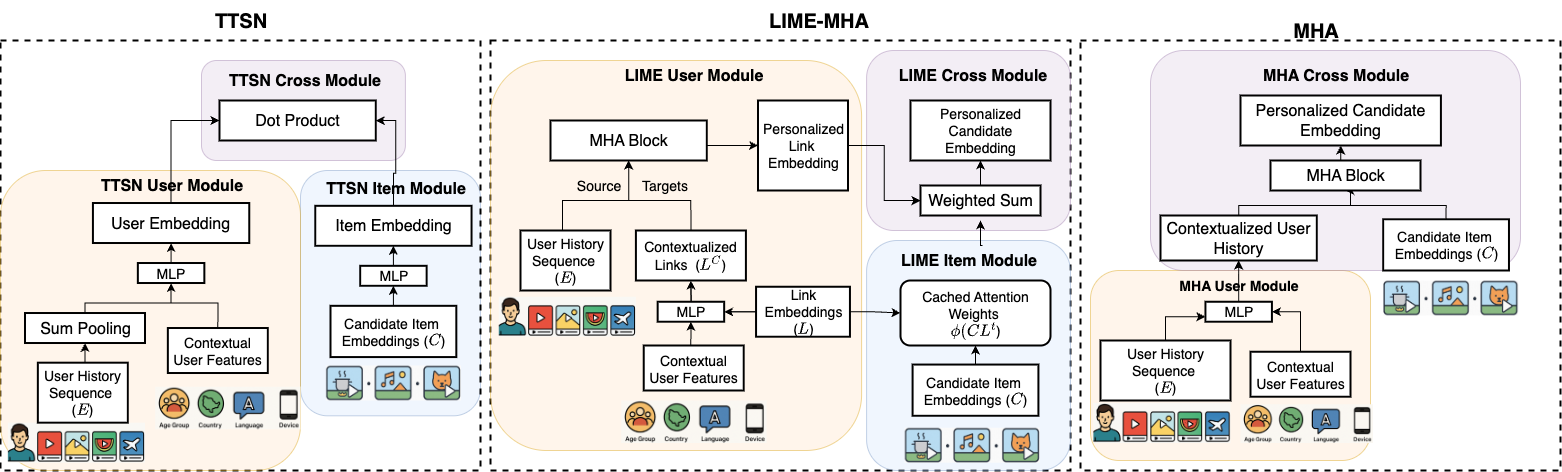}
% https://app.diagrams.net/#G1GeJXZ0wPWplhhRCV4VgcrG3cI9Q6Gceu#%7B%22pageId%22%3A%22pqgdSD683R3iJ5YSStIi%22%7D
%%\vspace{-55pt} % Reduce the space below this figure by 15 points

\caption{Architecture of TTSN (left), LIME-MHA (middle), and MHA (right). With a lightweight cross module using precomputed attention weights in a decoupled attention framework, LIME-MHA achieves MHA-level expressiveness with similar latency to TTSN.} \label{lime_train}
%\vspace{-0.5cm}
\end{figure*}

\subsection{From Two-Towers to LIME-MHA: Bridging Efficiency and Expressiveness}

While a two-tower model (TTSN) is highly efficient for scoring millions of items in retrieval, its expressivity is limited, as user-item interaction is confined to a late-stage dot product. At the other extreme, full cross-attention, prototyped by Multi-Head Attention (MHA), allows deep interaction between every candidate item and the entire user history. However, its computational cost, which scales with both candidate count ($M(U)$) and history length ($N(U)$), places the entire burden on the online interaction stage, making it prohibitively expensive.

LIME-MHA bridges this gap by introducing a small, fixed-size set of $\ell \ll N(U)$ auxiliary tokens, or \textbf{link embeddings} $L \in \mathbb{R}^{\ell \times d}$. These learned embeddings act as a compact summary of user interests, which are first personalized based on the user's history and then exposed to candidate items. This factorization is achieved via two MHA stages, effectively reducing the complexity from $O(M(U) \cdot N(U))$ to $O(\ell (M(U) + N(U)))$.

Multi-Head Attention (MHA), introduced by Vaswani et al.~\citep{vaswani2017attention}, is a function that maps $q$ query vectors $Q \in \mathbb{R}^{q \times d}$ to outputs using $n$ key and value vectors $K, V \in \mathbb{R}^{n \times d}$ as:
\begin{equation}
\label{eq:mha_block}
\mathrm{MHA}(Q, K, V; M; \theta) = \bigl(\phi(\tilde{Q} \tilde{K}^\top) \odot M \bigr) \tilde{V}
\end{equation}
where $\tilde{Q} = Q W_Q$, $\tilde{K} = K W_K$, $\tilde{V} = V W_V$, with learnable parameters $\theta = \{ W_Q, W_K, W_V \in \mathbb{R}^{d \times d} \}$. Here, $\phi$ is an activation function (e.g., scaled Softmax or SiLU) and $M \in \{0,1\}^{q \times n}$ is a binary mask. For intance, $J[i, j] :\equiv 1$ is the trivially all-1 mask pattern. $(A \odot B)_{ij} := A_{ij} B_{ij}$ stands for Hadamard product (elementwise multiplication).

The LIME-MHA architecture operates in two steps:

\paragraph{1. User-Side Link Personalization.} The globally shared link embeddings $L$ are first contextualized with user features $E^C$ (e.g., location, device type) via an MLP:
\begin{equation}
\label{eq:link_ctx}
    L^C = \mathrm{MLP}(L \oplus E^C)
\end{equation}
These contextualized links are then personalized by attending to the user's full interaction history $E$ using a single MHA layer, producing personalized link embeddings $L^P$:
\begin{equation} \label{eq:link-personalization_a}
    L^P = \text{MHA}(L^C, E, E; J; \theta)
\end{equation}
\paragraph{2. Candidate-Side Decoupled Interaction.} A key innovation of LIME is how candidate (target) embeddings $T$ interact with the personalized links. Instead of a standard MHA where keys and values both come from $L^P$, we use the raw, user-independent link embeddings $L$ as keys:
\begin{equation}
\label{eq:finalcand}
    O = \text{MHA}(T, L, L^P; J; \theta)=\phi(TL^\top)L^P
\end{equation}
This seemingly small change has a profound impact on efficiency. The attention weight matrix, $\phi(TL^\top) \in \mathbb{R}^{M(U) \times \ell}$, is now independent of the user. It can be pre-computed offline for all items in the corpus and cached. At inference time, this expensive matrix multiplication is replaced by a simple lookup, and the interaction reduces to a lightweight weighted sum of the personalized link embeddings $L^P$. This makes the serving latency per candidate effectively constant, $O(1)$, rather than scaling with history length.

\begin{figure*}[t]
% %\vspace{-2em}
\centering
%%\vspace{-65pt} % Reduce the space below this figure by 15 points
\includegraphics[width=1\textwidth]{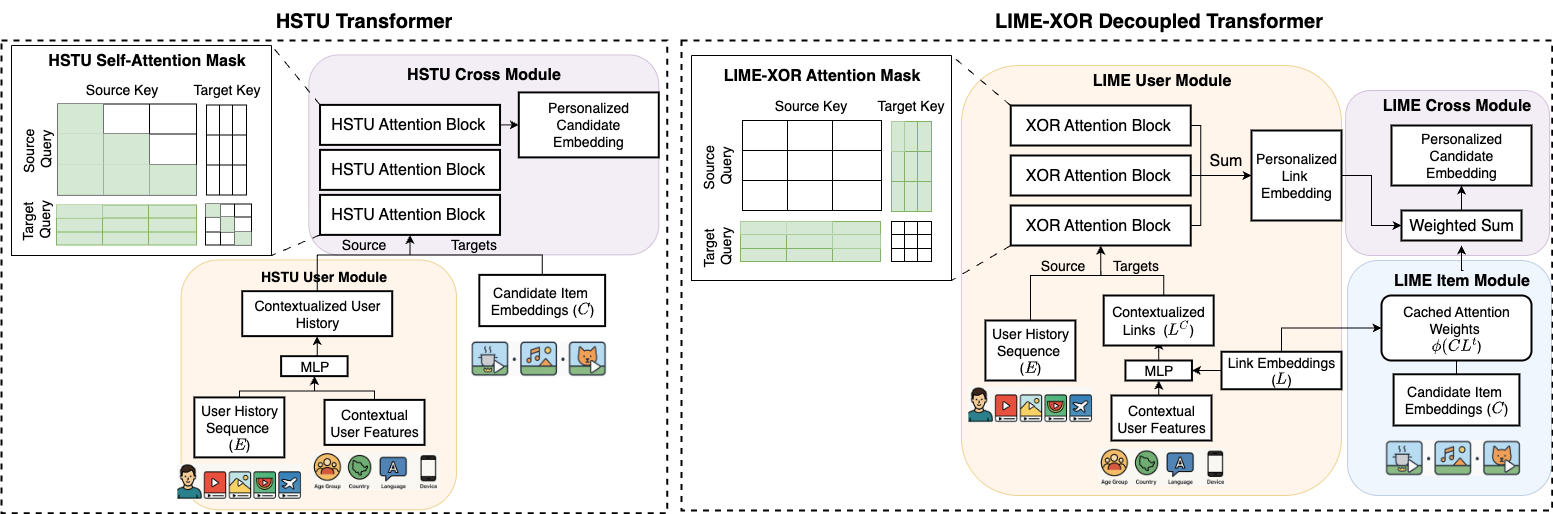}
% https://app.diagrams.net/#G1GeJXZ0wPWplhhRCV4VgcrG3cI9Q6Gceu#%7B%22pageId%22%3A%22pqgdSD683R3iJ5YSStIi%22%7D
%%\vspace{-55pt} % Reduce the space below this figure by 15 points

\caption{Architecture of HSTU Transformer (left) and LIME-XOR Decoupled Transformer (right) with a visual comparison of our proposed XOR attention mask against the standard HSTU causal self-attention mask.} \label{lime_xor_train}
%\vspace{-0.5cm}
\end{figure*}

\subsection{From LIME-MHA to LIME-XOR: Scaling to Deeper Interactions}

To enhance LIME's expressiveness, we can deepen the user-side module by stacking multiple interaction layers, akin to the architecture of a multi-layer Transformer. This computation is performed only once per user request, so its complexity does not impact the per-candidate scoring latency. A state-of-the-art approach for this would be to adapt a powerful sequential model block, such as the Hierarchical Sequential Transducer Unit (HSTU)~\citep{zhai2024actions}, which can be conceptually summarized as follows (see Appendix  A for full notations):
\begin{equation} \label{eq:hstu_brief}
\text{HSTUBlock}(X; \theta) = \text{GatedMLP}(\MHA(X, X, X; M_{\text{causal}}; \theta); \eta)
\end{equation}
where $X = E \oplus L^C$ is the concatenation of the user history and contextualized link embeddings. However, the standard HSTU attention uses causal self-attention mask, $M_\text{causal}[i, j] := i \geq j$, where every token in the user history attends to all preceding tokens (self-attention) and every candidate attends to the entire user history (cross-attention), results in a computational complexity of $O(N(U)^2 + N(U) \cdot M(U))$. This creates a major bottleneck for users with long interaction sequences or large candidate sets to rank.

To overcome this, we introduce XOR Attention (\textbf{XORA}), a novel attention kernel designed to replace the standard self-attention mechanism within the user-side module.
\begin{align}
\XORA(X, X, X; \theta) &= \MHA(X, X, X; M_{\text{xor}}; \theta) \label{eq:xora} \\
&= \MHA(E, L^C, L^C; J; \theta) \oplus \MHA(L^C, E, E; J; \theta)  \label{eq:xor_decomposition}
\end{align}
where $M_\text{xor}[i, j] := 1_{i \in [0, |E|)} \wedge 1_{j \in [0, |E|)}$ is the exclusive-or mask pattern that ensures the source and target embeddings attend to one another only. As depicted in Figure~3, the XOR mask structurally eliminates the expensive history-to-history ($E\leftrightarrow E$) interactions. Instead, it facilitates an efficient, two-way, block-wise attention \eqref{eq:xor_decomposition}: the link embeddings attend to the user history ($L^C \to E$), and crucially, the user history (source) embeddings now also attend back to the link embeddings ($E \to L^C$).

This modification provides two main advantages. First, it reduces the computational complexity from quadratic to linear, $O(\ell \cdot N(U))$, making deep, multi-layer processing of long histories feasible. Second, it enriches the model's expressivity by enabling the user history representation to be modulated by the global context of the link embeddings from the very first layer.

This leads to our advanced variant, \textbf{LIME-XOR}. In this model, we define an \textbf{XOR-Layer} by replacing the causal mask in \eqref{eq:hstu_brief} with our efficient XOR mask, $M_{\text{XOR}}$ in \eqref{eq:xora}. The personalized links are then computed by stacking and summing the outputs of $n$ such layers:
\begin{equation} \label{eq:link-personalization_b}
    L^P = \sum_{j=1}^{n} \text{GatedMLP}\left(\XORA\left(E \oplus L^C, E \oplus L^C, E \oplus L^C; \theta \right); \eta_j\right)
\end{equation}
This entire deep computation occurs on the user side, preserving the efficient, per-candidate scoring mechanism and LIME's overall scalability. Note that by using contextual links $L^C$ as targets instead of candidate embeddings $T$, the process remains fully decoupled from individual candidates during this stage.

\subsection{LIME's Architectural Advantages for Efficient Inference}

The LIME framework is designed from the ground up to resolve the conflict between model expressiveness and inference latency. Its core efficiency stems from a strategic decoupling of user-side and item-side computations, bridged by the link embeddings. This design enables a multi-stage inference pipeline that maximizes offline pre-computation and minimizes online work, making it highly scalable for real-world recommendation systems.

To facilitate this efficiency, we employ a technique called \textbf{decoupled attention}, used in the final candidate interaction stage (Equation \ref{eq:finalcand}). Instead of using the user-specific personalized links ($L^P$) for both keys and values, we use the raw, user-independent link embeddings ($L$) as keys and the personalized links ($L^P$) as values. This asymmetric structure ensures that the expensive Query-Key dot product, $\phi(CL^t)$, depends only on item-side information (candidate embeddings $C$ and raw links $L$). Consequently, this attention weight matrix can be pre-computed for all items in the corpus and stored in an efficient key-value store or index, such as FAISS~\citep{faiss}, effectively creating a cache of attention weights.

\begin{figure*}[t]
% %\vspace{-3em}
\centering
\includegraphics[width=1\textwidth]{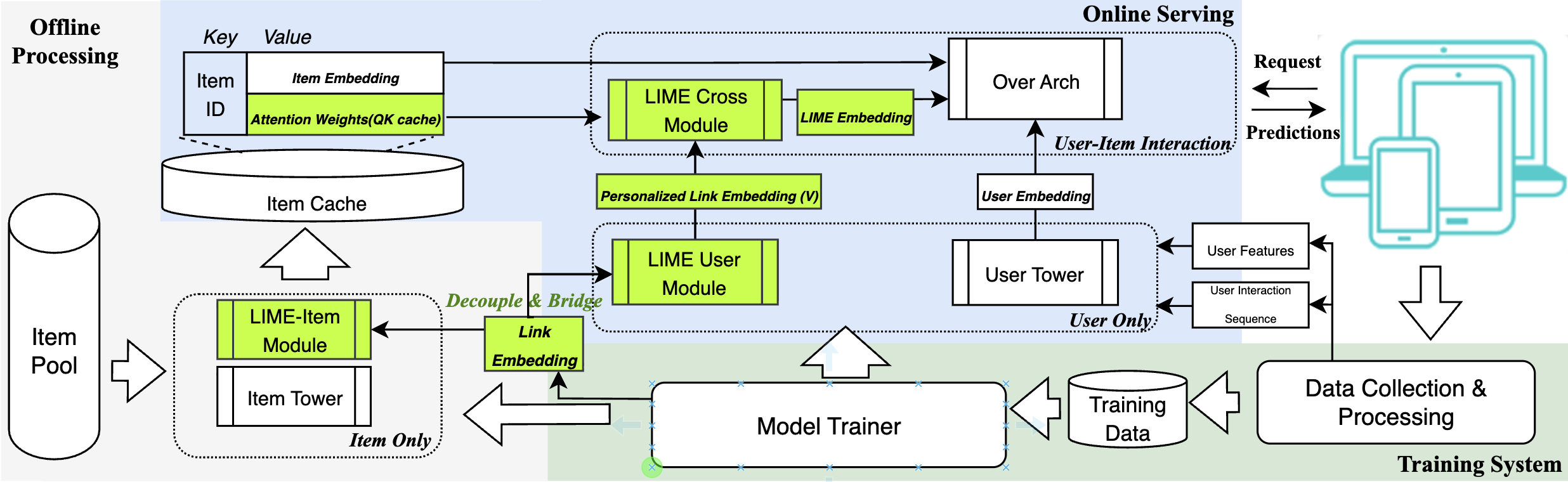}
% https://drive.google.com/file/d/1cFgKicCGtwvAGbd6Fk1EHPpyWKGM-Xqq/view?usp=sharing
\caption{LIME Inference as part of the entire recommendation system}\label{fig:lime_inference}
%\vspace{-0.5cm}
\end{figure*}

This architectural choice allows us to structure the entire inference process into three distinct stages, as illustrated in Figure~\ref{fig:lime_inference}.

\textbf{Offline Item-Side Pre-computation.} This stage is performed offline whenever the model or item is updated, eliminating redundant computation during online serving. Standard item features are processed by an item tower to produce item embeddings. The decoupled attention weights ($\phi(CL^\top)$) between all candidate item embeddings ($C$) and the raw link embeddings ($L$) are pre-computed and cached. This transforms the most intensive part of cross-attention into a simple lookup.

\textbf{Online User-Side Computation.} This stage runs once per user request and is independent of the number of candidates being scored. The user's context and interaction history are processed by the LIME user module (using either MHA or the multi-layer XOR Transformer) to produce the personalized link embeddings ($L^P$), as described in Equations~(\ref{eq:link_ctx})–(\ref{eq:link-personalization_b}).
Crucially, this user-side computation can be executed in parallel with the candidate retrieval process. In a production environment, its latency is therefore largely masked, making even a deep, multi-layer Transformer on the user side feasible.

\textbf{Lightweight User-Item Interaction.} This final stage is executed online for each candidate but is extremely lightweight. For each candidate, the pre-computed attention weights are retrieved from the QK Cache. These weights are used to perform a simple weighted sum over the personalized link embeddings ($L^P$) to generate the final LIME embedding. This embedding is then passed to a shallow interaction network for scoring.

By structuring inference this way, LIME achieves significant computational savings. Compared to a full cross-attention model like HSTU, which has a complexity of $\mathcal{O}(N(U) \cdot (N(U) + M(U)))$, LIME's complexity is reduced to $\mathcal{O}(\ell \cdot (N(U) + M(U)))$, where $\ell \ll N(U), M(U)$. More importantly, Model serving is reduced to a near-constant time operation with respect to the number of candidates, making LIME highly suitable for latency-sensitive ranking deployments with large candidate sets and long user histories.
% %\vspace{-1em}
\section{Experimental Results}

% \subsection{General Setup}

\textbf{Baselines and Skylines.}
We compare our approach against the highly efficient \textbf{TTSN} (Two-tower sparse network; Figure \ref{lime_train}) baseline, which sum-pools user history embeddings rather than using target attention. To establish performance upper-bounds (skylines), we evaluate two expressive but computationally prohibitive target-attention models: \textbf{MHA Skyline} (single-layer target attention) and \textbf{HSTU Skyline} (3 layers of causal self-attention and cross-attention between candidates and user history). For the details please refer Appendix~\ref{sec:experiment setup}.
% \textbf{Baselines.} We use \textbf{TTSN} (Two-tower sparse network): Depicted in Figure \ref{lime_train}, as the baseline. This baseline applies sum-pooling of all the user history embeddings instead of target attention with candidate embeddings.

% \textbf{Skylines.} We cannot launch naive target attention (candidate against user history) directly for long user histories and large numbers of candidates to rank. But such powerful models can serve as our skyline goal on performance. \textbf{MHA Skyline} computes a single layer of target attention between candidate items and user history items. \textbf{HSTU Skyline} computes 3 layers of causal self attention and cross attention at each layer, between the candidate items and user history items.

\textbf{LIME Variants.}
We evaluate three LIME variants differing in their link personalization modules (full details in Appendix X). \textbf{LIME-MHA} (Eq.~\eqref{eq:link-personalization_a}) uses standard attention to pool user history into link representations. \textbf{LIME-XOR} (Eq.~\eqref{eq:link-personalization_b}) deepens contextualization using a 3-layer HSTU with our efficient XOR mask. Finally, \textbf{LIME-XOR+Window} extends the XOR mask with a linear sliding-window pattern (attending to $\ell$ local neighbors, akin to Longformer \citep{beltagy2020longformer}) to jointly capture local and global interactions. To ensure a fair and controlled comparison, all other architectural components were held constant across all models (see details in Appendix~\ref{sec:exp_details}).
% \subsubsection{LIME Variants}
% To thoroughly evaluate our proposed architecture, we conducted experiments with two primary variants of LIME that differ specifically in the sophistication of the link personalization module.

% \textbf{LIME-MHA}: This is the fundamental implementation of our model, as described in Equation \eqref{eq:link-personalization_a}. It employs a standard attention mechanism to generate a personalized representation for each link by pooling information from the user's historical item embeddings.

% \textbf{LIME-XOR}: This advanced variant, detailed in Equation \eqref{eq:link-personalization_b}, incorporates 3-layers of HSTU between the user history and link embeddings for deeper contextualization with our proposed efficient XOR-style masking.

% \textbf{LIME-XOR+Window}: This variant extends the LIME-XOR attention mask by adding a sliding-window pattern over the user history, where each token attends to its $\ell$ neighbors on both sides (similar to Longformer \citep{beltagy2020longformer}). The attention remains linear: each token attends to $2 \ell$ tokens—$\ell$ link tokens and $\ell$ local neighbors. XOR attention captures global interactions between user history tokens, while the windowed attention captures local interactions.

\textbf{Metrics.} We present the normalized entropy metrics (NE) \citep{he2014practical} as well as (session) AUC metrics for all the variants and baselines/skylines. On the in-house industrial datasets, we measure accuracy on two tasks: video completion (VC) and watch time (WT). Normalized Entropy is defined for binary classification task as
% \begin{align*}
$\rm{NE}(\{(p_i, \ell_i)\}_{i=1}^n) := \frac{\sum_{i=1}^n \ell_i \log p_i + (1-\ell_i) \log (1 - p_i)}{\log (\sum_{i=1}^n \ell_i / n) }.$
% \end{align*}
Note that similar to logloss or binary cross entropy, lower NE means better accuracy. Usually $0.1\%$ drop in NE will lead to online metric improvement.

%\begin{table}[htbp]
%\small
%\setlength{\tabcolsep}{4pt} % Reduce horizontal padding
%\centering
%\caption{Statistics for public and industrial datasets.}
%\label{dataset_statistics}
%\begin{tabular}{lcc}
%\toprule
%Dataset & Max History Length & \# Interactions \\
%\midrule
%Taobao-Ad         & 50   & 25M \\
%KuaiRand          & 256  & 12M \\
%Industrial        & 1024 & $\sim$$10^{11}$ \\
%\bottomrule
%\end{tabular}
%\end{table}

\subsection{Industrial \& Public Experiments}

\textbf{Industry Experiments.} We take 3 days of logged data for training, and 6 hours for evaluation. Compared to the public datasets, this is a larger scale dataset with longer user history sequences. As shown in Table \ref{tab:merged_results}, both LIME variants significantly outperform the TTSN baseline. LIME-MHA nearly matches the MHA skyline, while the multi-layer LIME-XOR closes the gap further, achieving performance competitive with the much more complex HSTU skyline across all tasks, even with a 32x sequence compression rate. The 1.04\% VC NE improvement of LIME-XOR over the TTSN baseline is a significant gain, which translated to a +37.9\% VC and +28.6\% WT increase from LIME-ranked candidates during online A/B experiments (Table \ref{tab:merged_results}).

The online source-rate lifts in Table~\ref{tab:merged_results} are scoped to the source/generator traffic covered by the unified-preranker experiment and are not total platform-level lifts. In the same standard randomized experiment, the treatment group contained more than 10M users and the readout used a 7-day average. An engagement metric improved by $+0.111\%$ (90\% CI $[0.041\%,0.181\%]$), a watch-time metric by $+0.375\%$ (90\% CI $[0.205\%,0.545\%]$), and a session-level positive-engagement metric by $+0.233\%$ (90\% CI $[0.080\%,0.386\%]$). Broader watch-time and platform guardrails had no statistically significant regressions; capacity cost met the launch ROI bar.

% \subsection{Public Experiments Setup}
\textbf{Public Experiments.} We benchmark on public datasets, namely Taobao Ads \citep{lyu2020deep} and KuaiRand-1K dataset \citep{gao2022kuairand}. Taobao-Ads contains 25 million interactions with a maximum sequence length of 50 whereas KuaiRand-1K contains 12 million interactions with a maximum sequence length of 256. To unify data processing and evaluation framework, we leaned heavily on FuxiCTR, a comprehensive sequential recommendation model benchmark platform \citep{fuxictr}.
On both public datasets, LIME-MHA matches or outperforms the respective skyline with significant improvements over the sum-pooling baseline (Table \ref{tab:merged_results} (b)). The improvement is largest in the KuaiRand-1K dataset as it contains longer sequences (length 256). On KuaiRand-1K, LIME-XOR performs comparably to the HSTU skyline but slightly worse on Taobao-Ad, though still outperforming LIME-MHA models. This is likely due to shorter sequence lengths in Taobao-Ad, leading to noisier results.

We also benchmark existing sequence-compression methods, including Truncated MHA, which applies the MHA skyline only to the most recent $\ell$ interacted items, and Linformer-style LREA \citep{Song_2025} with a low rank of $\ell$. At the same low rank, LIME outperforms both compression techniques. Additionally we benchmark skyline models SASRec \citep{sasrec} (a state-of-the-art transformer) and DIN (Deep Interest Network) \citep{zhou2018deep} to validate our LIME variants and skylines are indeed improvements over existing state-of-the-art models.

To quantify variance and compare with the concurrent VISTA architecture, we additionally ran five seeds on the same Taobao-Ad split and training protocol. Table~\ref{tab:taobao_five_seed} reports mean and standard deviation at a matched budget of 32 VISTA seed embeddings or LIME links. LIME-XOR improves all three mean metrics over our VISTA implementation, and its AUC improvement over VISTA and TTSN is statistically significant. Across all methods the comparison remains metric-dependent: DIN has the best gAUC, HSTU the best AUC, and LIME-XOR the lowest log loss.

\begin{table}[t]
\centering
\caption{Five-seed Taobao-Ad results (mean $\pm$ standard deviation). Lower log loss is better.}
\label{tab:taobao_five_seed}
\small
\resizebox{\linewidth}{!}{%
\begin{tabular}{lccc}
\toprule
Model & gAUC & AUC & Log loss \\
\midrule
VISTA (32 seeds) & $0.569079 \pm 0.000465$ & $0.646075 \pm 0.000224$ & $0.192632 \pm 0.000106$ \\
TTSN & $0.570523 \pm 0.000328$ & $0.645368 \pm 0.000216$ & $0.192685 \pm 0.000227$ \\
DIN & $\mathbf{0.571175 \pm 0.000489}$ & $0.647357 \pm 0.000420$ & $0.192587 \pm 0.000458$ \\
HSTU & $0.570691 \pm 0.000500$ & $\mathbf{0.647669 \pm 0.000238}$ & $0.192597 \pm 0.000170$ \\
LIME-XOR (32 links) & $0.570567 \pm 0.000152$ & $0.646799 \pm 0.000239$ & $\mathbf{0.192545 \pm 0.000111}$ \\
\bottomrule
\end{tabular}%
}
\end{table}

\begin{comment}
\subsection{Ablation Studies}
In Figure \ref{fig:ablation_study}, we can see that 32 links seems to be an inflection point above which performance improvement is not witnessed and below which sequence compression may be too severe. Moreover, it is clear that modeling longer sequence lengths consistently brings larger wins with a fixed number of links with the eventual gaol being learning from lifelong user history sequences.
\end{comment}

% \begin{table}[htbp]
% %\small
% \centering
% \caption{Evaluation on public datasets.}
% \label{tab:combined_results}
% %\resizebox{\columnwidth}{!}{
% \begin{tabular}{l c | c}
% \toprule
% & \textbf{KuaiRand-1K} & \textbf{Taobao-Ad} \\
% Model & Click AUC & AUC  \\
% \midrule
% TTSN         & 0.7389 & 0.6452 \\
% DIN               & 0.7404 & 0.6468 \\
% SASRec            & 0.7419 & 0.6462\\
% \midrule
% Truncated MHA & 0.7351 & 0.6456 \\
% LREA & 0.7408 & 0.6447 \\
% \midrule
% LIME-MHA          & \textbf{0.7433} & \textbf{0.6465}\\ %\textbf{0.8843} \\
% MHA Skyline       & 0.7428 & 0.6464 \\
% \midrule
% %LIME              & \textbf{0.7452} & 0.6471 & 0.8845 \\
% LIME-XOR          & \textbf{0.7448} & 0.6467 \\
% HSTU Skyline      & 0.7444 & \textbf{0.6475} \\
% \bottomrule
% \end{tabular}
% %}
% \end{table}

\begin{table*}[t]
\centering
\caption{Offline, online, and public dataset experimental results.}
\label{tab:merged_results}
\small

% --- Left column: In-house results ---
\begin{minipage}[t]{0.49\linewidth}
    \centering
        \caption*{(a) Industrial Results}
    \resizebox{\linewidth}{!}{%
        \begin{tabular}{c|cc|cc}
        \toprule
        \multirow{2}{*}{Model} & \multicolumn{2}{c|}{\textbf{Offline}} & \multicolumn{2}{c}{\textbf{Online A/B}} \\
         & VC NE & WT AUC & VC & WT \\
        \midrule
        TTSN & -- & -- & -- & -- \\
        \midrule
        LIME-MHA & -0.72\%& +0.46\%& +28.6\% & +22.1\% \\
        MHA Sky. & \textbf{-0.73\%}& \textbf{+0.53\%}& N/A & N/A \\
        \midrule
        LIME-XOR & -1.04\%& +0.76\%& \textbf{+37.9\%} & \textbf{+28.6\%} \\
        LIME-XOR+Window & \textbf{-1.11\%} & \textbf{+0.80\%} & N/A & N/A \\
        HSTU Sky. & -1.06\% & +0.77\%& N/A & N/A \\
        \bottomrule
        \end{tabular}   }
    \caption*{MHA and HSTU Skylines cannot be tested in online setting due to high serving latency from the large number of candidates.}
\end{minipage}
\hfill
% --- Right column: Public results ---
\begin{minipage}[t]{0.49\linewidth}
    \centering
        \caption*{(b) Public Dataset Results}
    \resizebox{\linewidth}{!}{%
        \begin{tabular}{lcc}
        \toprule
        & \textbf{KuaiRand-1K} & \textbf{Taobao-Ad} \\
        Model & Click AUC & AUC \\
        \midrule
        TTSN & 0.7389 (+0\%)& 0.6452 (+0\%)\\
        DIN & 0.7404 (+0.20\%)& 0.6468 (+0.25\%)\\
        SASRec & 0.7419 (+0.41\%)& 0.6462 (+0.15\%)\\
        Trunc. MHA & 0.7351 (-0.51\%)& 0.6456  (+0.06\%)\\
        LREA & 0.7408 (+0.26\%)& 0.6447(-0.08\%)\\
        \midrule
        LIME-MHA & \textbf{0.7433} (+0.60\%)& \textbf{0.6465} (+0.20\%)\\
        MHA Sky. & 0.7428 (+0.53\%)& 0.6464 (+0.19\%)\\
        \midrule
        LIME-XOR & 0.7448 (+0.80\%)& 0.6467 (+0.23\%)\\
        LIME-XOR+Window & \textbf{0.7453} (+0.87\%)& 0.6468 (+0.25\%)\\
        HSTU Sky. & 0.7444 (+0.74\%)& \textbf{0.6475} (+0.36\%)\\
        \bottomrule
        \end{tabular}%
    }
\end{minipage}
\end{table*}

\begin{table}[t]
\centering
\caption{Controlled five-seed Taobao-Ad link/key ablation with 32 links (mean $\pm$ standard deviation). Lower log loss is better.}
\label{tab:link_key_ablation}
\small
\resizebox{\linewidth}{!}{%
\begin{tabular}{lcccc}
\toprule
Candidate-side basis & gAUC & AUC & Log loss & QK cacheable \\
\midrule
Learned global links & $0.570567 \pm 0.000152$ & $\mathbf{0.646799 \pm 0.000239}$ & $\mathbf{0.192545 \pm 0.000111}$ & Yes \\
Frozen random links & $0.570323 \pm 0.000255$ & $0.644434 \pm 0.000268$ & $0.192834 \pm 0.000243$ & Yes \\
Fixed item centroids & $\mathbf{0.571714 \pm 0.000337}$ & $0.644832 \pm 0.000288$ & $0.192608 \pm 0.000111$ & Yes \\
Personalized-link keys & $0.569040 \pm 0.000789$ & $0.643578 \pm 0.000553$ & $0.192915 \pm 0.000265$ & No \\
\bottomrule
\end{tabular}%
}
\end{table}

\begin{table*}[t]
\centering
\caption{Ablation Study and comparison to linear sparse attention variants on KuaiRand-1K dataset.}
\label{tab:ablation_results}
\small

% --- Left column ---
\begin{minipage}[t]{0.4\linewidth}
\centering
\caption*{(a) Ablation Study}
\resizebox{\linewidth}{!}{
\begin{tabular}{l c}
\toprule
\textbf{Model Variant} & \textbf{AUC} \\
\midrule
LIME-XOR & \textbf{0.7448} \\
LIME-XOR w/ Link Pooling & 0.7400 \\
LIME-XOR w/ Dot Product & 0.7404 \\
LIME-MLP & 0.7407 \\
LIME-XOR–MLP Hybrid & 0.7442 \\
\bottomrule
\end{tabular}
}
\end{minipage}
\hfill
% --- Right column ---
\begin{minipage}[t]{0.58\linewidth}
\centering
\caption*{(b) Comparison to SOTA Linear Attention Variants}
\resizebox{\linewidth}{!}{
\begin{tabular}{l c c}
\toprule
\textbf{Model Variant} & \textbf{AUC} & \textbf{Complexity} \\
\midrule
HSTU Skyline          & 0.7444 & $O(N(U)^2 + N(U) \cdot M(U))$ \\
\midrule
Linear Transformer    & 0.7423 & $O(N(U) \cdot M(U))$ \\
Longformer       & 0.7443 & $O(N(U) \cdot M(U))$ \\
LREA (Linformer)          & 0.7408 & $O(N(U) \cdot M(U))$ \\
\midrule
LIME-XOR              & 0.7448 & $O(N(U)\ell + M(U)\ell)$ \\
LIME-XOR+Window     & \textbf{0.7453} & $O(N(U)\ell + M(U)\ell)$ \\
\bottomrule
\end{tabular}
}
\end{minipage}
%\vspace{-0.5cm}
\end{table*}

\subsection{Ablation Study \& Scaling Laws}

We ablate various components of LIME-XOR to demonstrate the effectiveness of its design by considering the following variants. \textbf{LIME-XOR w/ Link Pooling}: Instead of decoupled attention with candidate embeddings, we apply sum pooling to the personalized link embeddings. \textbf{LIME-XOR w/ Dot Product}: Instead of decoupled attention with candidate embeddings, we pass the $\ell$ dot products between candidate embedding and personalized link embeddings to the final prediction MLP. \textbf{LIME-MLP}: Instead of leveraging attention mechanisms, we directly learn MLPs to generate personalized low-rank embeddings (after padding the user history to fixed length) and cached weights. \textbf{LIME-XOR-MLP}: We preserve the LIME-XOR user module with multi-layer XOR attention and replace the LIME item module cached attention weights with weights learned from an MLP on the candidate item embeddings.

The ablation results in Table~\ref{tab:ablation_results}(a) demonstrate that both multi-layer XOR attention from the LIME user module and the decoupled attention mechanism in the LIME item module are critical to strong performance. The attention mechanisms cannot easily be replaced by MLP, sum pooling, or dot products in either user or item module especially due to variable user history sequence lengths. The controlled link/key ablation in Table~\ref{tab:link_key_ablation} separately tests the candidate-side basis. Fixed item centroids have the best gAUC, while learned links retain the best AUC and log loss; the experiment therefore does not establish that learned anchors are uniformly superior to fixed alternatives. Personalized keys improve none of the three metrics and remove QK cacheability. For both NLP and CTR prediction tasks, transformer-based models (e.g., LLMs, HSTU) have shown improved performance as compute scales. On our large-scale ranking dataset, the decoupled transformer LIME-XOR similarly achieves substantial gains when scaling sequence length, link count, and model depth. Detailed results are provided in Appendix~\ref{sec:scaling law}.

\subsection{Sparse Linear Attention Variants}
We note that several alternate sparse linear attention mechanisms exist such as Linear Transformer \citep{katharopoulos2020transformers}, Longformer \citep{beltagy2020longformer}, and Linformer-style LREA \citep{wang2020linformer, Song_2025}. These can be integrated into the skyline transformer to reduce the quadratic self-attention cost ($O(N(U)^2)$) to linear $O(N(U)\ell)$ for low-rank $\ell$, though they will still suffer from $O(N(U) \cdot M(U))$ target attention cost. We can also integrate these linear attention variants within our proposed decoupled LIME framework to also optimize the target attention cost (e.g. LIME-XOR+Window).

From Figure~\ref{tab:ablation_results}(b), both Linear Transformer and Linformer-based methods have severe performance degradation but sliding-window based Longformer performs comparably to the HSTU skyline. However, it is still prohibitively expensive to serve (see Appendix \ref{sec:linear_perf} for latency analysis) due to the cross-attention cost. Yet, when we integrate sliding window attention into LIME-XOR we see improved performance with minimal latency, demonstrating the benefits of the LIME-XOR decoupled attention framework.

\subsection{Inference Speed}\label{sec:inference speed}
LIME is highly scalable for ranking large candidate sets, an advantage for pre-ranking and retrieval. Figure~\ref{fig:main_latency_analysis} demonstrates that while MHA and HSTU latency grows significantly with more candidates or longer user histories, LIME's latency remains nearly constant (see Appendix~\ref{sec: inference speed} for a detailed comparison). This robustness makes it suitable for settings with very long sequences (e.g.,$>$30k items). Furthermore, the user backbone computation can be parallelized with candidate retrieval, masking most of its latency in production and yielding even greater savings than shown in these benchmarks.

% One of LIME's primary advantages is its suitability for scoring extensive sets of ranking candidates, often numbering in the tens of thousands for candidate pre-ranking and retrieval (e.g., 10k). We present a performance benchmark in Figure ~\ref{fig:main_latency_analysis}, which measures inference latency on synthetic data as a function of user sequence length, number of candidates to rank, and embedding dimension $d$. A striking observation from this analysis is that while the inference latency of typical sequential models (e.g. MHA, HSTU) increases significantly with user history length and number of candidates, LIME's performance remains largely unaffected, showing only a minimal increase in latency. This highlights its robustness and scalability for early stage (coarse) ranking even in settings with very long user histories (e.g. 30k or longer) and large number of candidates to rank (e.g. 30k or more).

% It is worth noting that most of the inference latency can be masked in LIME variants by performing the user backbone in parallel with candidate retrieval. This is the dominating computational cost as it can involve multi-layer XOR Transformer whereas the lightweight interaction module is very inexpensive. Hence in production environments the inference latency savings are much more significant than the individual module benchmark results presented here.

% This tikzset is included here for completeness, but the user's query
% focuses on the content within the document body.
\tikzset{
  % Base style for all boxes
  box/.style={
    draw,                 % Draw a border
    fill=blue!10,         % Fill with a light blue color
    text centered,        % Center the text inside
    rounded corners=1pt,  % Sharper corners
    font=\sffamily\tiny,  % Use a very small sans-serif font
    inner sep=1.5pt       % Reduce padding inside the box
  },
  % Style for the large square boxes
  square/.style={
    box,
    minimum width=1.8cm,
    minimum height=1.8cm
  },
  % Style for the wide rectangular boxes
  wide/.style={
    box,
    minimum width=2.2cm,
    minimum height=0.8cm
  },
  % Style for the tall rectangular boxes
  tall/.style={
    box,
    minimum width=0.8cm,
    minimum height=2.2cm
  },
  % Style for the mathematical operators
  op/.style={
    font=\small,          % Make operators smaller
    node distance=0.15cm  % Reduce distance for operators
  }
}

\begin{figure}[htbp]
    \centering
    % --- LEFT PLOT ---
    \begin{subfigure}[b]{0.48\textwidth}
        \centering
        \begin{tikzpicture}
            \begin{axis}[
                width=\linewidth,
                height=5cm, % Fixed height for consistency
                xlabel={Singular Value Index},
                ylabel={Normalized Cumulative Singular Value},
                % Let pgfplots handle label positioning automatically
                label style={font=\footnotesize},
                tick label style={font=\footnotesize},
                xmin=-50, xmax=1050,
                ymin=0.45, ymax=1.1,
                ytick={0.5, 0.6, 0.7, 0.8, 0.9, 1.0},
                legend pos=south east,
                % Scaled down legend to prevent overlapping the axis lines
                legend style={font=\scriptsize, cells={align=left}},
            ]

            \addplot[blue, smooth, mark=*, mark options={scale=0.5, fill=blue}] coordinates {
                (0, 0.48) (10, 0.75) (25, 0.9) (50, 0.96) (100, 0.99) (200, 1.005) (400, 1.01) (1000, 1.015)
            };
            \addlegendentry{Self Attention Matrix}

            \addplot[red, smooth, mark=*, mark options={scale=0.5, fill=red}] coordinates {
                (0, 0.52) (10, 0.78) (25, 0.92) (50, 0.97) (100, 0.995) (200, 1.01) (400, 1.015) (1000, 1.02)
            };
            \addlegendentry{Cross Attention Matrix}

            % Vertical Dashed Line at x=50
            \draw[black, dashed] (axis cs:50, 0.45) -- (axis cs:50, 1.1);
            \end{axis}
        \end{tikzpicture}
    \end{subfigure}% <-- The '%' prevents unwanted whitespace
    \hfill
    % --- RIGHT PLOT ---
    \begin{subfigure}[b]{0.48\textwidth}
        \centering
        \begin{tikzpicture}
            \begin{axis}[
                width=\linewidth,
                height=5cm, % Match the height of the left plot
                xlabel={Singular Value Index},
                ylabel={Normalized Cumulative Singular Value},
                label style={font=\footnotesize},
                tick label style={font=\footnotesize},
                xmin=0, xmax=31,
                ymin=0, ymax=1.05,
                legend pos=south east,
                legend style={font=\scriptsize, cells={align=left}},
            ]

            \addplot[green!50!black, dashed, mark=triangle*, mark options={scale=0.8, fill=green!50!black}] coordinates {
                (1, 0.12) (5, 0.35) (10, 0.55) (15, 0.7) (20, 0.82) (25, 0.92) (30, 1.0)
            };
            \addlegendentry{Personalized Link Embeddings}

            \addplot[violet, dashdotted, mark=square*, mark options={scale=0.6, fill=violet}] coordinates {
               (1, 0.05) (5, 0.25) (10, 0.45) (15, 0.6) (20, 0.72) (25, 0.85) (30, 0.96)
            };
            \addlegendentry{Link Embeddings}

            \end{axis}
        \end{tikzpicture}
    \end{subfigure}
    %\vspace{-0.2cm}
    \caption{Left: cumulative singular value of self/cross attention matrices in a pretrained transformer model. Right: cumulative singular value of the raw/personalized link embeddings.}
    \label{fig:svd_analysis}
    %\vspace{-0.5cm}
\end{figure}
\section{Analysis and Discussions}
%\vspace{-0.1cm}

LIME effectively projects the high-dimensional user-user and user-candidate interaction spaces into low-rank subspaces, acting as a surrogate attention mechanism. This decomposition can be viewed as a structured approximation of the full attention matrix, analogous to techniques in sparse/dense low-rank compression (Figure~\ref{fig:attention_decomposition}).

To this end, we compute the singular value decomposition (SVD) of the self-attention and cross-attention matrices in a trained HSTU Transformer skyline model averaged across layers and heads over 256 sequence-candidate pairs where each sequence has length 1024 and we rank 1024 candidates against the sequence. In Figure \ref{fig:svd_analysis}, the results clearly demonstrate a long-tailed pattern where the largest 32 singular values (denoted by the vertical black line) capture more than 90\% of the information in both self and cross-attention matrices. We also compute the SVD of the raw link embeddings $L$ and personalized link embeddings $L^P$ of a trained LIME-XOR model and analyze the spectral distribution. The singular values of ${L}$ and $L^P$ demonstrate strong separation amongst the link embeddings with nearly full rank. This analysis indicates that 32 links are able to capture the majority of the information in the self-attention and cross-attention matrices with minimal redundancy for sequence lengths of 1024.

We also compare the factorized LIME-MHA attention distribution with direct MHA attention from independently trained models over 100k candidates and sampled users with history length 2,048. As the number of links increases from 1 to 128, Jensen--Shannon divergence decreases from 0.122 to 0.050 and cosine similarity increases from 0.837 to 0.923 (Table~\ref{tab:attention_approximation}). For context, uniform attention has Jensen--Shannon divergence 0.693. The 32-link model already reaches 0.055 divergence and 0.915 cosine similarity.

\begin{table}[t]
\centering
\caption{Similarity between factorized LIME-MHA and direct MHA attention distributions.}
\label{tab:attention_approximation}
\small
\begin{tabular}{rcc}
\toprule
Links & Jensen--Shannon divergence & Cosine similarity \\
\midrule
1 & 0.122 & 0.837 \\
4 & 0.072 & 0.902 \\
32 & 0.055 & 0.915 \\
128 & \textbf{0.050} & \textbf{0.923} \\
\bottomrule
\end{tabular}
\end{table}

%\vspace{-0.1cm}
\section{Conclusion}
%\vspace{-0.1cm}
We proposed LIME, a framework that resolves the efficiency-expressiveness trade-off in large-scale recommenders by using link embeddings as a low-rank approximation for target attention or more general transformer style sequence encoders. Experiments on public and industrial datasets show LIME matches the accuracy of state-of-the-art models like HSTU while drastically reducing inference latency, a result confirmed by significant online A/B test improvements. Our analysis confirmed that LIME's learned link embeddings effectively capture user interests, validating the low-rank hypothesis. By decoupling the user representation from target items, LIME offers a practical and powerful solution for building high-performance, scalable recommender systems.
\bibliographystyle{assets/plainnat}
\bibliography{bibliography}

\beginappendix
% \clearpage
\section{Limitations and Future Work}\label{sec:limitation}

Despite LIME's strong empirical performance, several limitations warrant discussion. First, LIME's effectiveness relies on the low-rank approximation assumption; domains with highly fragmented user interests may require higher-rank representations. Second, pre-computing QK caches requires periodic updates for evolving item catalogs, which makes it better suited for early stage ranking.

We are actively exploring the following directions: (1) building a larger pool of link embeddings and dynamically selecting link subsets based on user-side features or context, enabling personalized link allocation across different user segments,
% , (2) hierarchical link structures to capture interests at multiple granularities,
(2) extensions to multi-modal recommendation and cross-domain transfer learning to address the cold start user and item problem.
% , and (4) interpretability analysis to understand what semantic patterns individual links capture.

\section{Summary of Notations}

We summarize the key notations used in the paper in Table \ref{tab:lime-notations}.

\begin{table}[h!]
\centering
\caption{Notation Table}
\begin{tabular}{ll}
\hline
\textbf{Symbol} & \textbf{Description} \\
\hline
$U$ & User \\
$N(U)$ & Number of items in user $U$'s interaction history \\
$M(U)$ & Number of candidate items for user $U$ \\
\midrule
$E_j$ & Embedding representation of item $j$ \\
$E = \{E_j\}_{j=1}^{N(U)}$ & Sequence of user history embeddings, $E \in \mathbb{R}^{N(U) \times d}$ \\
$E^C$ & User context features (e.g., location, device type) \\
$T$ & Candidate (target) item embeddings \\
$d$ & Embedding dimension \\
$L \in \mathbb{R}^{\ell \times d}$ & Link embeddings (auxiliary tokens), $\ell \ll N(U)$ \\
$L^C$ & Contextualized link embeddings (personalized with user features) \\
$L^P$ & Personalized link embeddings (after attention to user history) \\
\midrule
$\mathrm{MHA}(Q, K, V; M; \theta)$ & Multi-Head Attention function \\
$W_Q, W_K, W_V$ & Learnable projection matrices for queries, keys, values ($\in \mathbb{R}^{d \times d}$) \\
$\theta, \eta$ & Set of learnable parameters \\
$\phi$ & Activation function (e.g., scaled Softmax, SiLU) \\
$O$ & Final LIME output embedding after candidate-side MHA \\
\midrule
$M$ & Binary mask for attention, $M \in \{0,1\}^{q \times n}$ \\
$J$ & All-ones mask pattern, $J[i,j] \equiv 1$ \\
$M_{\text{causal}}$ & Causal self-attention mask, $M_{\text{causal}}[i, j] := i \geq j$ \\
$M_{\text{xor}}$ & XOR attention mask, $M_{\text{xor}}[i, j] := 1_{i \in [0, |E|)} \wedge 1_{j \in [0, |E|)}$ \\
$\odot$ & Hadamard (elementwise) product \\
$\oplus$ & Concatenation operator \\
\midrule
$\text{GatedMLP}$ & MLP block with parameters $\eta$ to learn a gated element-wise product \\
$\text{HSTUBlock}(X; \theta)$ & Hierarchical Sequential Transducer Unit block with parameters $theta$ \\
$\XORA$ & XOR Attention Block \\
$n$ & Number of stacked XOR layers \\
\hline
\end{tabular}

\label{tab:lime-notations}
\end{table}

\section{Ranking Model Architecture Comparisons}
Outlined in Table \ref{tab:arch_comparison}, two-tower based models have the highest scalability in both candidate set size and user history length but suffer from extremely limited (e.g. dot product) late interaction. Conversely, transformer-based models have deep user-item interactions through full self-attention (amongst user history) and cross-attention (between candidates and user history) at every layer. However, transformer-based models suffer from high latency and low scalability on both user history length and candidate set size.

SIM \citep{pi2020search} and TWIN \citep{chang2023twin} improve scalability over Transformer-based methods through a two-stage approach. The general-search unit (GSU) first searches for the top-$K$ relevant user history interactions for a particular candidate and exact-search unit (ESU) performs a full cross-attention against the retrieved items. While the latency is reduced compared to Transformer, we also sacrifice some model expressivity when we completely remove user history self-attention and the GSU runtime complexity is $O(MN)$ which can still be inefficient.

On the other hand, LIME is able to model deep user-item interactions (through the decoupled Transformer framework introduced in Section 3) and achieves high scalability to both candidate set size and user history length when $L \ll N$.

\begin{table}[htbp]

\centering
\caption{Comparison of Ranking Model Architectures where $N$ is the user history length, $M$ is the number of candidates to rank, and $L \ll N$ is the LIME-compressed history size.}
\label{tab:arch_comparison}
\resizebox{0.9\linewidth}{!}{%
\begin{tabular}{l|l|l|l|l}
\hline
\textbf{Axis} & \textbf{Transformers (HSTU)} & \textbf{SIM/TWIN} & \textbf{Two-Tower} & \textbf{LIME (Ours)}  \\
\hline \hline
User-Item Interaction     & Deep & Medium                     & Limited late interaction & Deep \\
Latency         & High   & Medium             & Low                 & Low  \\
Complexity    & $O(M N + N^2)$   & $O(MN)$        & $O(N)$          &$O(M\ell + N\ell)$ \\
Pre-computation           & Minimal    & Minimal              & Item Emb         & QK Attention Weights \\
\hline
\end{tabular}
}
\end{table}

\section{Experiment Implementation Details}\label{sec:exp_details}

\subsection{Baseline Setup}\label{sec:experiment setup}

\subsubsection{Baselines}
We test multiple baseline models ranging from the simplest sum-pooling of user history embeddings to other sequence compression techniques listed below.

\textbf{TTSN} (Two-tower sparse network): Depicted in Figure \ref{lime_train}, this baseline applies sum-pooling of all the user history embeddings instead of target attention with candidate embeddings.

\subsubsection{Skylines}
We cannot launch naive target attention (candidate against user history) directly for long user histories and large numbers of candidates to rank. But such powerful models can serve as our skyline goal on performance.

\textbf{MHA Skyline} computes a single layer of target attention between candidate items and user history items.

\textbf{HSTU Skyline} computes 3 layers of causal self attention and cross attention at each layer, between the candidate items and user history items.

\subsubsection{LIME Variants}
To thoroughly evaluate our proposed architecture, we conducted experiments with two primary variants of LIME that differ specifically in the sophistication of the link personalization module.

\textbf{LIME-MHA}: This is the fundamental implementation of our model, as described in Equation \eqref{eq:link-personalization_a}. It employs a standard attention mechanism to generate a personalized representation for each link by pooling information from the user's historical item embeddings.

\textbf{LIME-XOR}: This advanced variant, detailed in Equation \eqref{eq:link-personalization_b}, incorporates 3-layers of HSTU between the user history and link embeddings for deeper contextualization with our proposed efficient XOR-style masking.

\textbf{LIME-XOR+Window}: This variant extends the LIME-XOR attention mask by adding a sliding-window pattern over the user history, where each token attends to its $\ell$ neighbors on both sides (similar to Longformer \citep{beltagy2020longformer}). The attention remains linear: each token attends to $2 \ell$ tokens—$\ell$ link tokens and $\ell$ local neighbors. XOR attention captures global interactions between user history tokens, while the windowed attention captures local interactions.

\subsection{Dataset Preparation}
For the Taobao Ad dataset, we leverage the preprocessed version provided by the FuxiCTR \citep{fuxictr} sequential modeling platform. For the KuaiRand dataset \citep{gao2022kuairand}, we preprocess it ourselves by partitioning the first 14 days of interactions for training and last 2 days for testing. We also discarded all items from the train and test set with less than 30 total interactions for more reliable results. For the industrial dataset, we took 3 days of logged data for model training and 6 hours of data for evaluation.

\subsubsection{Model Hyperparameters}
For both public datasets and the industrial datasets, we fix model hyperparameters and seeds for all variants for a fair comparison (see Table \ref{tab:model_params}). The industrial dataset is of the largest scale both in terms of the number of interactions and maximum sequence length. Due to the large size of the industrial dataset, we only train for a single epoch in a streaming single-pass setting for the industrial dataset.

\begin{table}[h]
\centering
\caption{Model Hyperparameters for KuaiRand, TaoBao Ad, and Industrial datasets}
\label{tab:model_params}
\begin{tabular}{lccc}
\hline
\textbf{Parameter} & \textbf{KuaiRand} & \textbf{TaoBao Ad} & \textbf{Industrial} \\
\hline
Num. of Interactions & $12 \times 10^6$ & $25 \times 10^6$ & $100 \times 10^9$ \\
Learning Rate       & $1 \times 10^{-4}$      & $1 \times 10^{-3}$      & $4 \times 10^{-4}$      \\
Batch Size          & 1024                   & 8192                   & 1024                   \\
Number of Heads     & 4                      & 4                      & 4                      \\
Epochs              & 2                      & 10                     & 1                      \\
Embedding Dimension & 32                     & 32                     & 256                    \\
Number of Links     & 16                     & 8                      & 32                     \\
Max Sequence Length & 256                    & 50                     & 1024                    \\
Interaction MLP     & [512, 128, 64]          & [512, 256, 128]        & [96]                 \\
\hline
\end{tabular}
\end{table}

\subsection{Model Design}
Several additional design choices were crucial to stabilize training and ensure generalization:

\begin{itemize}
    \item \textbf{Normalization:} We apply Layer normalization before each linear projection in the QKV projections. Without normalization, the embeddings can drift toward high magnitudes, which can collapse attention weights.
    \item \textbf{Link Initialization:} The raw link embeddings $\ell_i$ are initialized with samples from a standard normal distribution. This encourages diversity and allows the model to discover interpretable interest clusters during training.
    \item \textbf{Attention Function:} In single-layer MHA experiments we use scaled Softmax which performs the best and in multi-layer transformer experiments we leverage SiLU as the attention activation $\phi$.
    \item \textbf{Training Objective:} We use a standard cross-entropy loss with multi-task objectives per request. Other loss terms (e.g., contrastive objectives or auxiliary disentanglement losses) may be explored in future work.
\end{itemize}

\section{Low Rank Expressions for Self-attention and Cross-attention}
LIME effectively projects the high-dimensional user-user and user-candidate interaction spaces into low-rank subspaces, acting as a surrogate attention mechanism:
\begin{align*}
\mathrm{MHA}(T, E) &\approx \mathrm{MHA}(T, L) \cdot \mathrm{MHA}(L, E), \\
\mathrm{MHA}(E, E) &\approx \mathrm{MHA}(E, L) \cdot \mathrm{MHA}(L, E).
\end{align*}
This decomposition can be viewed as a structured approximation of the full attention matrix, analogous to techniques in sparse/dense low-rank compression (e.g. Performer, Linformer, SVDNet).

\section{Information Bottleneck Perspective}
From an information-theoretic viewpoint, LIME compresses the user history $E$ into a set of link embeddings $L$, which are optimized to retain relevance to the user’s behavior (via personalized attention pooling) and retain discriminative power for candidate ($T$) ranking.

This fits into the Information Bottleneck (IB) principle:
\begin{align*}
\min_{{L}} \mathcal{I}(L;E) - \beta \mathcal{I}(L;T)
\end{align*}
where $\mathcal{I}$ denotes mutual information. That is, we retain only those aspects of the user history that are useful for predicting interaction with candidates.

\begin{figure}[h!]
    \centering
    % --- TOP DIAGRAM ---
    \begin{tikzpicture}[
        node distance=0.4cm and 0.6cm,
        every node/.style={font=\small\sffamily, align=center},
        square/.style={draw, rectangle, minimum width=2.5cm, minimum height=1cm, rounded corners=3pt},
        tall/.style={draw, rectangle, minimum width=1cm, minimum height=2.2cm, rounded corners=3pt},
        wide/.style={draw, rectangle, minimum width=3cm, minimum height=1cm, rounded corners=3pt},
        op/.style={font=\large\sffamily}
    ]

    \node[square] (mha1) {MHA\\(Q = Candidates,\\ KV = UIH)};
    \node[op, right=of mha1] (approx1) {$\approx$};
    \node[square, right=of approx1] (mha2) {MHA\\(Q = Candidates,\\ KV = Links)};
    \node[op, right=of mha2] (times) {$\times$};
    \node[tall, right=of times] (mha3) {\rotatebox{90}{\shortstack{MHA\\(Q = Links,\\ KV = UIH)}}};
    \end{tikzpicture}

    %\vspace{0cm} % vertical space between diagrams

    % --- BOTTOM DIAGRAM ---
    \begin{tikzpicture}[
        node distance=0.4cm and 0.6cm,
        every node/.style={font=\small\sffamily, align=center},
        square/.style={draw, rectangle, minimum width=2.5cm, minimum height=1cm, rounded corners=3pt},
        tall/.style={draw, rectangle, minimum width=1cm, minimum height=2.2cm, rounded corners=3pt},
        wide/.style={draw, rectangle, minimum width=3cm, minimum height=1cm, rounded corners=3pt},
        op/.style={font=\large\sffamily}
    ]

    \node[square] (mha4) {MHA\\(QKV = UIH)};
    \node[op, right=of mha4] (approx2) {$\approx$};
    \node[tall, right=of approx2] (mha5) {\rotatebox{90}{\shortstack{MHA\\(Q = UIH,\\ KV = Links)}}};
    \node[op, right=of mha5] (plus) {$\times$};
    \node[square, right=of plus] (mha6) {MHA\\(Q = Links,\\ KV = UIH)};
    \end{tikzpicture}

    \caption{Top: Cross-attention as a low-rank product. Bottom: Self-attention as two low-rank cross-attentions. UIH stands for user interaction history.}
    \label{fig:attention_decomposition}
\end{figure}

\section{Inference Speed Comparison}\label{sec: inference speed}
Depicted in Figure \ref{fig:main_latency_analysis} and Table \ref{tab:latency_comparison_vertical}, we observe that LIME-MHA and LIME-XOR have significantly lower latencies under large candidate sets to rank, long user histories, and high QK dimensions (due to the attention weight cacheing).

\begin{table*}[htbp]
    \centering
    \caption{Latency (in seconds) comparison for different models under varying conditions. The columns correspond to the models: LIME-MHA, MHA Skyline, HSTU Skyline, and LIME-XOR. All numerical values are truncated to two decimal places.}
    \label{tab:latency_comparison_vertical}

    % Global settings for siunitx: truncate numbers to 2 decimal places
    \sisetup{
        round-mode      = places,
        round-precision = 2,
    }

    % --- TABLE 1: vs. Number of Candidates ---
    \begin{subtable}{\textwidth}
        \centering
        \caption{vs. Number of Candidates (FlashAttention V2)}
        \label{tab:vs_candidates}
        \begin{tabular}{ r|S|S|S|S }
            \toprule
            {\# Cand.} & {LIME-MHA} & {MHA Sky.} & {HSTU Sky.} & {LIME-XOR} \\
            \midrule
16    & 0.84 & 0.87 & 3.60 & 2.62 \\
32    & 0.83 & 0.87 & 3.66 & 2.53 \\
64    & 0.88 & 0.88 & 3.74 & 2.52 \\
128   & 0.84 & 0.87 & 3.64 & 2.55 \\
256   & 0.87 & 0.88 & 3.67 & 2.52 \\
512   & 0.85 & 0.88 & 4.01 & 2.56 \\
1024  & 0.87 & 0.89 & 3.76 & 2.61 \\
2048  & 0.85 & 0.95 & 3.70 & 2.43 \\
4096  & 0.83 & 1.06 & 4.17 & 2.51 \\
8192  & 0.86 & 1.11 & 4.42 & 2.53 \\
16384 & 0.87 & 1.64 & 7.86 & 3.21 \\
32768 & 0.99 & 2.76 & 11.66 & 4.06 \\
65536 & 1.35 & 4.25 & 23.83 & 5.95 \\
            \bottomrule
        \end{tabular}
    \end{subtable}

    %\vspace{1em}

    % --- TABLE 2: vs. User History Length ---
    \begin{subtable}{\textwidth}
        \centering
        \caption{vs. User History Length (FlashAttention V2)}
        \label{tab:vs_history}
        \begin{tabular}{ r|S|S|S|S }
            \toprule
            {Hist. Len.} & {LIME-MHA} & {MHA Sky.} & {HSTU Sky.} & {LIME-XOR} \\
            \midrule
16    & 0.68 & 0.72 & 4.18 & 3.02 \\
32    & 0.76 & 0.80 & 3.97 & 3.00 \\
64    & 0.83 & 0.72 & 4.19 & 3.01 \\
128   & 0.71 & 0.71 & 5.75 & 2.97 \\
256   & 0.72 & 0.70 & 4.00 & 3.25 \\
512   & 0.81 & 0.73 & 4.17 & 3.29 \\
1024  & 0.77 & 0.85 & 4.11 & 2.94 \\
2048  & 0.68 & 0.91 & 4.83 & 3.05 \\
4096  & 0.87 & 1.63 & 7.83 & 3.04 \\
8192  & 0.88 & 1.62 & 7.72 & 3.06 \\
16384 & 1.39 & 3.56 & 14.00 & 3.84 \\
32768 & 2.54 & 7.29 & 32.78 & 6.25 \\
            \bottomrule
        \end{tabular}
    \end{subtable}

    %\vspace{1em}

    % --- TABLE 3: vs. QK Dimension ---
    \begin{subtable}{\textwidth}
        \centering
        \caption{vs. QK Dimension (PyTorch)}
        \label{tab:vs_qkdim}
        \begin{tabular}{ r|S|S|S|S }
            \toprule
            {QK dim} & {LIME-MHA} & {MHA Sky.} & {HSTU Sky.} & {LIME-XOR} \\
            \midrule
            16   & 1.22 & 1.28 & 7.46 & 2.56 \\
            32   & 1.28 & 1.38 & 7.10 & 2.55 \\
            64   & 1.25 & 1.31 & 7.02 & 2.57 \\
            128  & 1.26 & 1.44 & 7.21 & 2.59 \\
            256  & 1.26 & 1.88 & 8.18 & 2.65 \\
            512  & 1.24 & 1.66 & 7.54 & 2.64 \\
            1024 & 1.30 & 3.13 & 11.02 & 2.90 \\
            2048 & 1.51 & 3.31 & 10.78 & 3.36 \\
            4096 & 3.07 & 13.57 & 35.23 & 7.93 \\
            \bottomrule
        \end{tabular}
    \end{subtable}

\end{table*}

\paragraph{Candidate-side cache decomposition.}
We separately benchmark the 32-link candidate-side operator with batch size 1, history length 50, dimension 96, four heads, 20 warmups, and 100 CUDA-event repetitions on an NVIDIA PG509-210 Ampere GPU. Table~\ref{tab:cache_decomposition} includes lookup and weighted aggregation rather than reporting lookup alone. BF16 caching yields a $1.04$--$1.39\times$ P50 speedup over the uncached global-key operator in this eager module-level benchmark because aggregation remains the dominant cost. Eager int8 dequantization removes this gain at 100k candidates. At 10k candidates, BF16 lookup, weighted aggregation, and their combined execution have P50/P95 latencies of 0.038/0.054, 0.117/0.178, and 0.162/0.234 ms, respectively.

\begin{table*}[t]
\centering
\caption{P50 candidate-side latency in milliseconds for 32 links. Cached totals include lookup and weighted aggregation.}
\label{tab:cache_decomposition}
\small
\resizebox{\textwidth}{!}{%
\begin{tabular}{rccccccc}
\toprule
Candidates & Uncached global & BF16 lookup & Weighted sum & BF16 total & Int8 lookup+dequant & Int8 total & Personalized keys \\
\midrule
100 & 0.198 & 0.037 & 0.119 & 0.157 & 0.068 & 0.194 & 0.193 \\
1,000 & 0.202 & 0.037 & 0.124 & 0.146 & 0.068 & 0.192 & 0.205 \\
10,000 & 0.191 & 0.038 & 0.117 & 0.162 & 0.070 & 0.186 & 0.196 \\
100,000 & 0.313 & 0.092 & 0.222 & 0.301 & 0.207 & 0.414 & 0.321 \\
\bottomrule
\end{tabular}%
}
\end{table*}

For a catalog $\mathcal I$, $\ell$ links, $H$ heads, and $b$ bytes per cached score, the current per-head layout occupies $|\mathcal I|\ell Hb$ bytes. Thus 100M items, 32 links, four heads, and int8 scores require 12.8 GB; 3.2 GB would apply only to a head-aggregated layout. The current cache is rebuilt in model memory at each model publication (every five minutes), so lookup requires no network request. The observed miss rate is below 0.1\%, and a staleness test changed online NE by less than 0.1\% after several hours. Full production cache-build cost, miss-fallback latency, and deployed hardware/layout remain to be measured.

\section{Scaling Law for LIME-XOR Decoupled Transformer}\label{sec:scaling law}
Transformers (such as HSTU) have demonstrated strong scaling law for large-scale recommendation systems, mirroring scaling laws from LLMs from NLP. For our decoupled transformer, LIME-XOR, we test scaling across three axes on the large-scale industrial dataset: sequence length, number of links, and number of layers.

From Table \ref{link_seq_scale}, we observe that scaling sequence length is very effective in improving both VC NE and WT AUC. However, it is only effective under the presence of sufficient links. Scaling sequence length from 2k to 4k under 32 links is neutral but similar scaling with 256 links demonstrates significant NE and AUC wins. Moreover, for shorter sequences (e.g. 1k) scaling the number of links seems to reach an inflection point faster than scaling links for longer sequences.

From Table \ref{tab:layer_scale}, we observe that scaling number of layers demonstrates significant NE and AUC improvements which helps demonstrate that LIME-XOR can also achieve scaling law up to the limits we were able to test.

\begin{table}[h]
\centering

\begin{subtable}[t]{\textwidth}
\centering
\begin{tabular}{c|cccc}
\toprule
         & 32     & 64      & 128     & 256     \\
\midrule
1k       & 0\%    & -0.04\% & -0.19\% & -0.20\% \\
2k       & -0.16\% & -0.19\% & -0.34\% & -0.39\% \\
4k       & -0.11\% & -0.36\% & -0.36\% & \textbf{-0.55\%} \\
\bottomrule
\end{tabular}
\caption{VC NE \% improvement across sequence lengths and link counts}
\end{subtable}

%\vspace{1em} % optional spacing between subtables

\begin{subtable}[t]{\textwidth}
\centering
\begin{tabular}{c|cccc}
\toprule
         & 32     & 64      & 128     & 256     \\
\midrule
1k       & 0\%    & +0.01\% & +0.08\% & +0.11\% \\
2k       & +0.03\% & +0.05\% & +0.18\% & +0.18\% \\
4k       & +0.03\% & +0.13\% & +0.17\% & \textbf{+0.27\%} \\
\bottomrule
\end{tabular}
\caption{WT AUC across sequence lengths and link counts}
\end{subtable}

\caption{Performance across varying sequence lengths (1k, 2k, 4k) and link counts (32, 64, 128, 256). Each cell reports metric value and relative change from the baseline (1k, 32 links).}
\label{link_seq_scale}
\end{table}

Table~\ref{tab:link_latency_scaling} complements the quality scaling results with candidate-side latency and cache storage. At 10k candidates the eager kernels are largely launch-bound, so latency is not monotonic in $\ell$, while storage grows linearly.

\begin{table*}[t]
\centering
\caption{P50 latency at 10k candidates and estimated int8 cache size for 100M items and four heads.}
\label{tab:link_latency_scaling}
\small
\resizebox{\textwidth}{!}{%
\begin{tabular}{rccccc}
\toprule
Links & User encoding (ms) & Uncached global (ms) & BF16 cached total (ms) & Int8 cached total (ms) & Int8 cache (GB) \\
\midrule
8 & 0.330 & 0.187 & \textbf{0.143} & 0.186 & 3.2 \\
16 & 0.347 & 0.193 & \textbf{0.154} & 0.182 & 6.4 \\
32 & 0.340 & 0.191 & \textbf{0.162} & 0.186 & 12.8 \\
64 & 0.349 & 0.195 & \textbf{0.151} & 0.186 & 25.6 \\
128 & 0.476 & 0.212 & \textbf{0.157} & 0.197 & 51.2 \\
\bottomrule
\end{tabular}%
}
\end{table*}

On the industrial dataset, we also compare LIME with geometric alternatives. Relative to LIME (NE change 0\%, latency 1.39 ms), two-round and four-round LSH regress NE by $+0.50\%$ and $+0.43\%$ at 3.31 and 6.41 ms, respectively. Clustered attention with 128 and 1,024 centroids regresses NE by $+0.32\%$ and $+0.19\%$ at 1.72 and 2.85 ms. Positive NE changes are worse.

\begin{table}[h]
\centering
\begin{tabular}{l|c|c}
\toprule
Number of Layers & VC NE \% Improvement & WT AUC \% Improvement \\
\midrule
3  & 0\%    & 0\%    \\
6  & -0.25\% & +0.12\% \\
9  & -0.34\% & +0.18\% \\
12 & \textbf{-0.38\%} & \textbf{+0.19\%} \\
\bottomrule
\end{tabular}
\caption{VC NE and WT AUC for increasing model depth on fixed 1k sequence length with 32 links.}
\label{tab:layer_scale}
\end{table}

\section{Derivations for XOR Attention Kernel}
\subsection{Forward Pass}
Let the output of the XOR-attention be $O = (O[S], O[T])$, where $S, T$ stand for source and target respectively. In the context of LIME, source is the user history item embeddings, while target is the link embeddings. Similarly define $Q[S], Q[T], K[S], K[T], V[S], V[T]$ to be the source and target portion of the query, key, value embedding sequences.
\begin{align*}
O[T] &= \phi(Q[T]K[S]^\top) V[S] \\
O[S] &= \phi(Q[S]K[T]^\top) V[T]
\end{align*}
\subsection{Backward Pass}
Let $dO[T]$ denote an infinitesimally small change in $O[T]$, also known as the its differential. Similarly define $dO[S], dV[S], dV[T], dQ[S], dQ[T], dK[S], dK[T]$.

Trivially we have
\begin{align*}
\frac{\partial O[T]}{\partial Q[S]} &= \frac{\partial O[T]}{\partial K[T]} = \frac{\partial O[T]}{\partial V[T]} = 0 \\
\frac{\partial O[S]}{\partial Q[T]} &= \frac{\partial O[S]}{\partial K[S]} = \frac{\partial O[S]}{\partial V[S]} = 0
\end{align*}
The total differential of the loss function is given by
\begin{align*}
dL = \tr \left( \left(\frac{\partial L}{\partial O[S]} \right)^\top dO[S]\right) + \tr \left( \left(\frac{\partial L}{\partial O[T]} \right)^\top dO[T]\right).
\end{align*}

Since $dO[S] = \phi(Q[S]K[T]^\top) dV[T]$ and $dO[T] = \phi(Q[T]K[S]^\top) dV[S]$,
\begin{align*}
dL_V = \tr\left(\left(\frac{\partial L}{\partial O[S]}\right)^\top \phi(Q[S]K[T]^\top) dV[T]\right) + \tr\left(\left(\frac{\partial L}{\partial O[T]}\right)^\top \phi(Q[T]K[S]^\top) dV[S]\right).
\end{align*}
By comparing coefficient with the matrix chain rule $dL_{V[T]} = \tr\left(\left(\frac{\partial L}{\partial V[T]}\right)^\top dV[T]\right)$, we see that
\begin{align*}
\frac{\partial L}{\partial V[T]} = \phi(K[T]Q[S]^\top) \frac{\partial L}{\partial O[S]}.
\end{align*}
Similarly
\begin{align*}
\frac{\partial L}{\partial V[S]} = \phi(K[S]Q[T]^\top) \frac{\partial L}{\partial O[T]}.
\end{align*}
Next for derivatives with respect to $Q$, we have
\begin{align*}
dL_{Q[S]} &= \tr\left(\left(\frac{\partial L}{\partial O[S]}\right)^\top \phi'(Q[S]K[T]^\top) V[T]K[T]^\top dQ[S]\right) \\
dL_{Q[T]} &= \tr\left(\left(\frac{\partial L}{\partial O[T]}\right)^\top \phi'(Q[T]K[S]^\top) V[S]K[S]^\top dQ[T]\right).
\end{align*}
Hence
\begin{align*}
\frac{\partial L}{\partial Q[S]} &= K[T]V[T]^\top \phi'(K[T] Q[S]^\top) \frac{\partial L}{\partial O[S]} \\
\frac{\partial L}{\partial Q[T]} &= K[S]V[S]^\top \phi'(K[S] Q[T]^\top) \frac{\partial L}{\partial O[T]}.
\end{align*}
Finally from
\begin{align*}
dL_{K[T]} &= \tr\left(\left(\frac{\partial L}{\partial O[S]}\right)^\top \phi'(Q[S]K[T]^\top) V[T]Q[S]^\top dK[T]\right) \\
dL_{K[S]} &= \tr\left(\left(\frac{\partial L}{\partial O[T]}\right)^\top \phi'(Q[T]K[S]^\top) V[S] Q[T]^\top dK[S]\right),
\end{align*}
we get
\begin{align*}
\frac{\partial L}{\partial K[S]} &= Q[T]V[S]^\top \phi'(K[S]Q[T]^\top) \frac{\partial L}{\partial O[T]} \\
\frac{\partial L}{\partial K[T]} &= Q[S]V[T]^\top \phi'(K[T]Q[S]^\top) \frac{\partial L}{\partial O[S]}
\end{align*}

\section{Triton PseudoCode}
\begin{algorithm}[H]

\caption{XOR Mask and Denominator Computation}
\label{alg:xor-mask}
\begin{algorithmic}[1]
\Require Query index $i$, key index $j$, number of sources $n_s$
\Ensure Binary mask $m$, normalization denominator $d$
\State $\text{is\_src}_q \gets (i < n_s)$
\State $\text{is\_src}_k \gets (j < n_s)$
\State $m \gets \text{is\_src}_q \oplus \text{is\_src}_k$ \Comment{Attend iff exactly one is source}
\State $d \gets \text{is\_src}_q \ ?\ n_t : n_s$ \Comment{Normalize by opposite partition size}

\State \Return $m, d$
\end{algorithmic}

\end{algorithm}

\begin{algorithm}[H]

\caption{Forward Pass with Block Range Selection}
\label{alg:forward}
\begin{algorithmic}[1]
\Require $\mathbf{Q}, \mathbf{K}, \mathbf{V} \in \mathbb{R}^{n \times d}$, number of sources $n_s$
\Ensure Output $\mathbf{O} \in \mathbb{R}^{n \times d}$
% --- FIX: Renamed loop variable (no subscript) ---
\For{$q_\text{start} \gets 0$ to $n$ step $B_M$} \Comment{Parallel over query blocks}
    \State $\mathbf{Q}_b \gets \mathbf{Q}[q_\text{start} : q_\text{start}+B_M, :]$ \Comment{Load to SRAM}
    \State $\text{acc} \gets \mathbf{0}_{B_M \times d}$
    \State
    \State \textbf{// Block range selection (critical optimization)}
    \If{$q_\text{start} + B_M \leq n_s$} \Comment{Pure source queries}
        \State $k_{\text{lo}}, k_{\text{hi}} \gets n_s, n$ \Comment{Load target keys only}
    \ElsIf{$q_\text{start} \geq n_s$} \Comment{Pure target queries}
        \State $k_{\text{lo}}, k_{\text{hi}} \gets 0, n_s$ \Comment{Load source keys only}
    \Else \Comment{Boundary case}
        \State $k_{\text{lo}}, k_{\text{hi}} \gets 0, n$
    \EndIf
    \State
    % --- FIX: Renamed loop variable (no subscript) ---
    \For{$k_\text{start} \gets k_{\text{lo}}$ to $k_{\text{hi}}$ step $B_N$} \Comment{Over selected K,V blocks}
        \State $\mathbf{K}_b \gets \mathbf{K}[:, k_\text{start} : k_\text{start}+B_N]$ \Comment{Load K,V to SRAM}
        \State $\mathbf{V}_b \gets \mathbf{V}[k_\text{start} : k_\text{start}+B_N, :]$
        \State $\mathbf{S} \gets \mathbf{Q}_b \mathbf{K}_b$ \Comment{Compute scores (tensor cores)}
        \State

        % --- FIX: Correctly nested loops ---
        \For{$i \gets 0$ to $B_M$} \Comment{Mask \& normalize in registers}
            \For{$j \gets 0$ to $B_N$}
                \State $m, d \gets \textsc{XorMask}(q_\text{start}+i, k_\text{start}+j, n_s)$
                \State $\mathbf{S}[i,j] \gets m \ ?\ \text{SiLU}(\mathbf{S}[i,j]) / d : 0$
            \EndFor
        \EndFor
        \State
        \State $\text{acc} \gets \text{acc} + \mathbf{S} \mathbf{V}_b$ \Comment{Accumulate (tensor cores)}
    \EndFor
    \State
    \State $\mathbf{O}[q_\text{start} : q_\text{start}+B_M, :] \gets \text{acc}$ \Comment{Write to HBM}
\EndFor

\State \Return $\mathbf{O}$
\end{algorithmic}

\end{algorithm}

\begin{algorithm}[H]

\caption{Backward Pass (Transposed Access Pattern)}
\label{alg:backward}
\begin{algorithmic}[1]
\Require $\mathbf{dO}, \mathbf{Q}, \mathbf{K}, \mathbf{V}$, number of sources $n_s$
\Ensure Gradients $\mathbf{dQ}, \mathbf{dK}, \mathbf{dV}$
\State $\mathbf{dQ}, \mathbf{dK}, \mathbf{dV} \gets \mathbf{0}$
\For{$k_{\text{start}} \gets 0$ to $n$ step $B_N$} \Comment{Parallel over K,V blocks}
    \State $\mathbf{K}_b \gets \mathbf{K}[:, k_{\text{start}} : k_{\text{start}}+B_N]$ \Comment{Load K,V to SRAM (resident)}
    \State $\mathbf{V}_b \gets \mathbf{V}[k_{\text{start}} : k_{\text{start}}+B_N, :]$
    \State $\mathbf{dK}_{\text{acc}}, \mathbf{dV}_{\text{acc}} \gets \mathbf{0}_{d \times B_N}, \mathbf{0}_{B_N \times d}$
    \State
    \If{$k_{\text{start}} + B_N \leq n_s$} \Comment{Source K,V block}
        \State $q_{\text{lo}}, q_{\text{hi}} \gets n_s, n$ \Comment{Process target queries only}
    \ElsIf{$k_{\text{start}} \geq n_s$} \Comment{Target K,V block}
        \State $q_{\text{lo}}, q_{\text{hi}} \gets 0, n_s$ \Comment{Process source queries only}
    \Else \Comment{Boundary case}
        \State $q_{\text{lo}}, q_{\text{hi}} \gets 0, n$
    \EndIf
    \State
    \For{$q_{\text{start}} \gets q_{\text{lo}}$ to $q_{\text{hi}}$ step $B_M$} \Comment{Over selected Q blocks}
        \State $\mathbf{Q}_b \gets \mathbf{Q}[q_{\text{start}} : q_{\text{start}}+B_M, :]$
        \State $\mathbf{dO}_b \gets \mathbf{dO}[q_{\text{start}} : q_{\text{start}}+B_M, :]$
        \State $\mathbf{S} \gets \mathbf{Q}_b \mathbf{K}_b$ \Comment{Recompute forward (activation remat)}
        \State
        \For{$i \gets 0$ to $B_M$, $j \gets 0$ to $B_N$}
            \State $m, d \gets \textsc{XorMask}(q_{\text{start}}+i, k_{\text{start}}+j, n_s)$
            \If{$m$}
                \State $s_{ij} \gets \text{SiLU}(\mathbf{S}[i,j]) / d$
                \State $\mathbf{S}[i,j] \gets s_{ij} \cdot (1 - s_{ij}) \cdot (1 + \mathbf{S}[i,j]) / d$ \Comment{SiLU gradient}
            \Else
                \State $\mathbf{S}[i,j] \gets 0$
            \EndIf
        \EndFor
        \State
        \State $\mathbf{dV}_{\text{acc}} \gets \mathbf{dV}_{\text{acc}} + \mathbf{S}^\top \mathbf{dO}_b$ \Comment{Accumulate (tensor cores)}
        \State $\mathbf{dK}_{\text{acc}} \gets \mathbf{dK}_{\text{acc}} + (\mathbf{S}^\top \mathbf{Q}_b)^\top$
        \State \textsc{AtomicAdd}($\mathbf{dQ}[q_{\text{start}} : q_{\text{start}}+B_M, :]$, $(\mathbf{dO}_b \mathbf{V}_b^\top) \mathbf{S}^\top$)
    \EndFor
    \State
    \State $\mathbf{dK}[:, k_{\text{start}} : k_{\text{start}}+B_N] \gets \mathbf{dK}_{\text{acc}}$ \Comment{Write gradients}
    \State $\mathbf{dV}[k_{\text{start}} : k_{\text{start}}+B_N, :] \gets \mathbf{dV}_{\text{acc}}$
\EndFor

\State \Return $\mathbf{dQ}, \mathbf{dK}, \mathbf{dV}$
\end{algorithmic}

\end{algorithm}

\section{Linear Attention Performance \& Latency Analysis} \label{sec:linear_perf}
In Figure \ref{tab:lin_attn_variants}, we compare various linear attention methods (not decoupled so still more expensive than LIME) to both LIME variants and skyline HSTU. The results demonstrate that LIME variants consistently match (or exceed) HSTU skyline AUC under varying sequence length and QK-dimension settings.

We benchmark all variants using PyTorch's FlexAttention library \citep{dong2024flex} which enables us to flexibly compare various masking strategies, ensuring that unnecessary blocks of computation are skipped, without writing custom Triton logic for each type of linear attention. In Figure \ref{tab:linear_benchmark}, we observe that Longformer is less expensive than Skyline HSTU when scaling user history length and number of candidates. However, it is still more expensive than LIME decoupled variants in complexity ($O(NM)$ vs $O(N\ell + M\ell)$). We observe it is more than twice as expensive across all user history lengths and spikes at user histories longer than 4096 and at number of candidates to rank greater than 2048.

\begin{figure}[t]
\centering

\begin{minipage}[t]{0.98\linewidth}
\centering
\caption{AUC Comparison of Sparse Attention Variants with columns reporting on a particular Sequence Length/QK Dimension (e.g. 256/32 means 256 sequence length, 32 QK-dimension).}
\label{tab:lin_attn_variants}
\begin{tabular}{
    l
    >{\centering\arraybackslash}p{1.6cm}
    >{\centering\arraybackslash}p{1.6cm}
    >{\centering\arraybackslash}p{1.6cm}
    >{\centering\arraybackslash}p{1.6cm}
    >{\centering\arraybackslash}p{2.4cm}
}
\toprule
\textbf{Variant} &
\(\textbf{256/128}\) &
\(\textbf{64/128}\) &
\(\textbf{256/32}\) &
\(\textbf{64/32}\) &
\textbf{Complexity} \\
\midrule

HSTU Skyline
    & \(0.7444\) & \(0.7300\) & \(0.7425\) & \(\mathbf{0.7296}\)
    & {\small \(\begin{aligned}
        O(N^2 + NM)
    \end{aligned}\)} \\

\midrule

Linear Trans.
    & \(0.7423\) & \(0.7292\) & \(0.7404\) & \(0.7278\)
    & {\small \(\,O(NM)\)} \\

Longformer
    & \(0.7443\) & \(0.7300\) & \(0.7425\) & \(0.7292\)
    & {\small \(\,O(NM)\)} \\

Linformer (LREA)
    & \(0.7408\) & \(0.7289\) & \(0.7406\) & \(0.7275\)
    & {\small \(\,O(NM)\)} \\

\midrule

LIME-XOR
    & \(0.7448\) & \(0.7301\) & \(\mathbf{0.7427}\) & \(0.7295\)
    & {\small \(\begin{aligned}
        O(N\ell + M\ell)
    \end{aligned}\)} \\

LIME-XOR+Win.
    & \(\mathbf{0.7453}\) & \(\mathbf{0.7305}\) & \(0.7425\) & \(0.7292\)
    & {\small \(\begin{aligned}
        O(N\ell + M\ell)
    \end{aligned}\)} \\

\bottomrule
\end{tabular}
\end{minipage}

\end{figure}

\begin{figure}[!htbp]
\centering

% ============================================================
% (a) FIRST ROW LEFT
% ============================================================
\begin{subfigure}[b]{0.49\linewidth}
\centering
% --- PLOT A ---
\begin{tikzpicture}
\pgfplotstableread{
max_uih_len   flex_xor_lime_stu   flex_window   flex_causal   flex_xor_window
32      7.415149   12.730281    32.301508    8.040612
64      7.253107   12.748127    31.937929    7.214512
128     7.319783   12.745853    32.069983    7.335038
256     7.320938   13.158813    32.876239    7.153169
512     7.312041   13.390839    34.449177    7.177161
1024    7.603478   15.509830    37.659830    7.135627
2048    7.228394   15.897291    44.816231    7.724213
4096    9.050267   22.809677    60.596692   10.481513
8192    14.982631  45.130901    99.664714   17.760276
16384   26.995037  126.608619   205.544113   32.804021
}\datatableA

\begin{semilogxaxis}[
xlabel={User History Length},
ylabel={Latency (ms)},
grid=both,
grid style={dashed, gray!30},
width=\linewidth,
height=5.5cm,
log basis x=2,
xtick={32,128,512,2048,8192,16384},
xticklabel style={rotate=45, anchor=east},
legend=false
]
\addplot[color=blue, mark=*] table[x=max_uih_len, y=flex_xor_lime_stu]{\datatableA};
\addplot[color=orange, mark=diamond*] table[x=max_uih_len, y=flex_xor_window]{\datatableA};
\addplot[color=green!50!black, mark=triangle*] table[x=max_uih_len, y=flex_window]{\datatableA};
\addplot[color=red, mark=x] table[x=max_uih_len, y=flex_causal]{\datatableA};
\end{semilogxaxis}
\end{tikzpicture}
\end{subfigure}%
% ← IMPORTANT: NO SPACE OR LINE BREAK HERE!
\hfill%
% ============================================================
% (b) FIRST ROW RIGHT
% ============================================================
\begin{subfigure}[b]{0.49\linewidth}
\centering
% --- PLOT B ---
\begin{tikzpicture}
\pgfplotstableread{
max_nro_len flex_xor_lime_stu flex_window flex_causal flex_xor_window
32      10.844105   12.338676   31.789882   11.204267
64      10.990144   12.177311   31.325211   11.814336
128     10.806154   12.102628   31.172334   11.158446
256     10.953121   12.523176   32.343293   11.083847
512     10.813794   12.787020   33.861469   11.024693
1024    10.496813   16.234868   36.362219   11.808427
2048    10.441883   21.182938   43.202838   11.076778
4096    10.979415   34.813191   58.427507   11.100659
8192    11.915411   47.167072   95.375485   12.084083
16384   13.738379   83.026995  197.619699   14.836940
}\datatableB

\begin{semilogxaxis}[
xlabel={Number of Candidates},
grid=both,
grid style={dashed, gray!30},
width=\linewidth,
height=5.5cm,
log basis x=2,
xtick={32,128,512,2048,8192,16384},
xticklabel style={rotate=45, anchor=east},
legend=false
]
\addplot[color=blue, mark=*] table[x=max_nro_len, y=flex_xor_lime_stu]{\datatableB};
\addplot[color=orange, mark=diamond*] table[x=max_nro_len, y=flex_xor_window]{\datatableB};
\addplot[color=green!50!black, mark=triangle*] table[x=max_nro_len, y=flex_window]{\datatableB};
\addplot[color=red, mark=x] table[x=max_nro_len, y=flex_causal]{\datatableB};
\end{semilogxaxis}
\end{tikzpicture}
\end{subfigure}

% ============================================================
% (c) SECOND ROW
% ============================================================
%\vspace{0.45cm}

\begin{subfigure}[b]{0.7\linewidth}
\centering
% --- PLOT C ---
\begin{tikzpicture}
\pgfplotstableread{
D  flex_xor_lime_stu  flex_window  flex_causal  flex_xor_window
32    5.857475   9.875852   9.803458   5.944529
64    6.235035   9.901905   9.585288   6.163326
128   15.407384  32.813745  31.133903  16.108340
256   16.688295  34.007845  32.182095  17.984858
512   16.812439  38.178252  35.515504  18.479813
1024  17.272563  45.227912  43.208220  16.717705
2048  27.902156  77.721407  70.459572  20.523340
4096  46.632256 159.779989 163.909678  47.503308
}\datatableC

\begin{semilogxaxis}[
xlabel={QK Embedding Dim},
ylabel={Latency (ms)},
grid=both,
grid style={dashed, gray!30},
width=\linewidth,
height=5.5cm,
log basis x=2,
xtick={32,64,128,256,512,1024,2048,4096},
xticklabel style={rotate=45, anchor=east},
legend=false
]
\addplot[color=blue, mark=*] table[x=D, y=flex_xor_lime_stu]{\datatableC};
\addplot[color=orange, mark=diamond*] table[x=D, y=flex_xor_window]{\datatableC};
\addplot[color=green!50!black, mark=triangle*] table[x=D, y=flex_window]{\datatableC};
\addplot[color=red, mark=x] table[x=D, y=flex_causal]{\datatableC};
\end{semilogxaxis}
\end{tikzpicture}
\end{subfigure}

% ============================================================
% SHARED LEGEND
% ============================================================
%\vspace{0.25cm}

\begin{tikzpicture}
\begin{axis}[
    hide axis,
    xmin=0, xmax=10, ymin=0, ymax=1,
    legend columns=-1,
    legend style={draw=none, column sep=0.35cm, anchor=center, font=\small},
]
\addlegendimage{color=blue, mark=*}
\addlegendentry{XOR-LIME}
\addlegendimage{color=orange, mark=diamond*}
\addlegendentry{XOR+Window}
\addlegendimage{color=green!50!black, mark=triangle*}
\addlegendentry{Longformer}
\addlegendimage{color=red, mark=x}
\addlegendentry{HSTU}
\end{axis}
\end{tikzpicture}

\caption{
Latency comparison of linear attention to LIME-XOR and HSTU.}
\label{tab:linear_benchmark}
\end{figure}

\section{Link Clustering Analysis}
\label{sec:link_clustering}
For one of the LIME-XOR models trained on the industrial dataset, we pick several links and
examine the items that have the largest attention weights with that particular link. In doing so,
we find that the clustered items for each link share strong semantic properties (e.g., video topic,
length, etc.). Some examples include:

\begin{itemize}
    \item \textbf{Link 2}: Long videos
    \item \textbf{Link 3}: Food \& Cooking videos
    \item \textbf{Link 5}: Dog \& Animal videos
    \item \textbf{Link 10}: Furniture \& DIY videos
    \item \textbf{Link 30}: Sports videos
    \item \textbf{Link 31}: Foreign videos
\end{itemize}

\section{LIME Performance on Cold-Start Users}
Although LIME is not explicitly designed for cold-start users, we observe that it delivers topline improvements for this group in online A/B testing that are comparable to those seen in the overall user cohort. To further understand this empirical behavior, we evaluate our pretrained HSTU skyline and LIME-XOR models on the cold-start subset of KuaiRand-1k. In this evaluation (see Table \ref{tab:cold_start_eval}), we vary the maximum number of prior interactions a user may have to qualify as a cold-start user.

\begin{table}[h!]
\centering
\caption{LIME-XOR vs HSTU Performance (AUC) on the cold-start user cohort in KuaiRand-1k}
\begin{tabular}{lcc}
\hline
Cold-Start User Threshold & LIME-XOR & HSTU \\
\hline
$N(U) \le 4$  & \textbf{0.7022} & 0.7016 \\
$N(U) \le 8$  & \textbf{0.7179} & 0.7173 \\
$N(U) \le 16$ & \textbf{0.7284} & 0.7273 \\
$N(U) \le 32$ & \textbf{0.7369} & 0.7356 \\
\hline
\end{tabular}
\label{tab:cold_start_eval}

\end{table}

One hypothesis for why LIME leads to improved performance for cold-start users is that LIME compresses a user's history into a fixed length embedding (equal to number of links) regardless of whether the user is a power user or cold-start user. This normalization allows the model to extract useful patterns even from sparse histories, reducing variance and improving robustness compared to models like HSTU that depend more directly on the number of prior interactions.

\section{LLM Usage in Preparation}\label{sec:llm usage}
We used LLM significantly to polish the language of all sections in the paper, including the abstract and appendix sections. The tikz codes in Figure~\ref{fig:main_latency_analysis}, ~\ref{fig:svd_analysis}, ~\ref{fig:attention_decomposition}, and \ref{tab:linear_benchmark}  were generated based on data points we collected ourselves. Most of the tables were also formatted by LLM.
%%%%%%%%%%%%%%%%%%%%%%%%%%%%%%%%%%%%%%%%%%%%%%%%%%%%%%%%%%%%
\section{Broader Impacts} \label{sec:Broader Impacts}

Our work on the Link-based User-Item Interaction Modeling for Efficient inference framework presents several potential positive and negative societal impacts.
\paragraph{Positive Societal Impacts.} LIME achieves state-of-the-art cross-attention expressiveness while operating within the strict efficiency budget of a two-tower model. The architecture delivers a 10x inference speedup over computationally intensive ranking models. When deployed on major platforms serving billions of users, this efficiency drastically reduces the computational resources required for large-scale recommendation systems. Besides, our empirical evaluations demonstrate that LIME delivers topline performance improvements for "cold-start" users who have limited prior interaction histories. By compressing a user's history into a fixed-length embedding regardless of their activity level, the model reliably extracts useful patterns from sparse data and reduces variance.

\paragraph{Potential Negative Impacts.} Because LIME is highly effective at deep, contextual personalization based on long user histories , it carries the inherent risk of over-optimizing for user engagement. Highly accurate sequential recommendation models can inadvertently reinforce echo chambers and limit the overall diversity of content a user is exposed to.
\end{document}